\documentclass[preprint,12pt]{revtex4}
\usepackage{graphicx}
\usepackage{amsmath}
\usepackage{amssymb}
\usepackage{amsfonts}
\usepackage{upgreek}
\usepackage{txfonts}
\usepackage{comment}

\usepackage{hyperref}
\usepackage{xcolor}
\definecolor{dark-red}{rgb}{0.4,0.15,0.15}
\definecolor{dark-blue}{rgb}{0.15,0.15,0.4}
\definecolor{medium-blue}{rgb}{0,0,0.5}
\hypersetup{
    colorlinks, linkcolor={dark-blue},
    citecolor={dark-blue}, urlcolor={medium-blue}
}

\begin{document}

\title{Quantum kinetics of ultracold fermions coupled to an optical resonator}

\author{Francesco Piazza$^1$}
\author{Philipp Strack$^{2,3}$}
\affiliation{$^{1}$Physik Department, Technische Universit\"at M\"unchen, 85747 Garching, Germany}
\affiliation{$^{2}$ Department of Physics, Harvard University, Cambridge MA 02138}
\affiliation{$^{3}$ Institut f\"ur Theoretische Physik, Universit\"at zu K\"oln, D-50937 Cologne, Germany}

\date{\today}

\begin{abstract}
We study the far-from-equilibrium statistical mechanics of 
periodically driven fermionic atoms in a lossy optical resonator. We show that the interplay of the 
Fermi surface with cavity losses leads to sub-natural 
cavity linewidth narrowing, squeezed light, and out-of-equilibrium quantum statistics of the atoms. 
Adapting the Keldysh approach, we set-up and solve a quantum kinetic Boltzmann equation 
in a systematic $1/N$ expansion with $N$ the number of atoms. In the strict 
thermodynamic limit $N,V\rightarrow \infty$, $N/V=\text{const.}$ we find the atoms (fermions or bosons) 
remain immune against cavity-induced 
heating or cooling. At next-to-leading order in $1/N$, we find a ``one-way thermalization'' of the atoms 
determined by cavity decay.
We argue that, in absence of an equilibrium fluctuation-dissipation relation, 
the long-time limit $\Delta t \rightarrow \infty$ does not commute with the thermodynamic limit $N\rightarrow \infty$, 
such that for the physically relevant case of large but finite $N$, the dynamics  
ultimately becomes strongly coupled, especially close to the superradiance phase transition. 
%
\end{abstract}

\maketitle
\tableofcontents

\section{Introduction}

Interacting light-matter systems that couple confined electromagnetic fields with 
ultracold atoms or qubits are emerging as an appealing research 
area combining physics from condensed matter, quantum optics, and out-of-equilibrium statistical mechanics.
Recent experiments in cavity quantum electrodynamics (QED) \cite{rbh_qed_rmp,kimble_1998, mabuchi_2002, haroche_book} 
have begun to scale up the prototypical situation of a single qubit coupled to a single photon to many atoms \cite{vuletic_2003,courteille_2007,eth_2007,reichel_2007,st_kurn_2007,eth_2010,barrett_2012,hemmerich_2012,zimmermann_2013,hemmerich_2014,cavity_rmp} and many photon modes \cite{sarang_2010,philipp_2011}.
%
%
\begin{figure} [b]
\includegraphics[width=140mm]{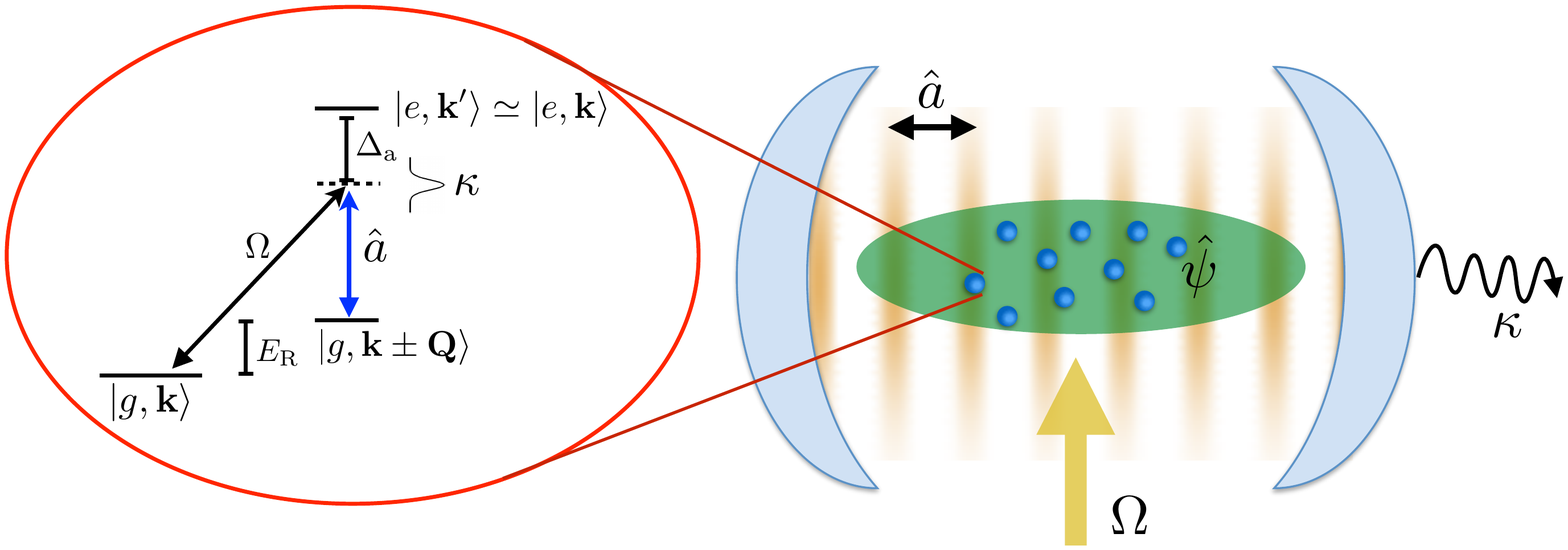}
\caption{Illustration of $N$ non-interacting atoms with two internal electronic levels trapped inside a cavity.
The atoms are driven periodically by a pump laser and scatter photons from the pump laser into the cavity, 
and dissipate into the environment with decay rate $\kappa$.}
\label{model_system}
\end{figure}

A paradigmatic manifestation of the collective behavior in such systems is the existence of a superradiant self-organization transition \cite{ritsch_2002} with coherently driven atoms, already observed with a thermal cloud \cite{vuletic_2003,barrett_2012} and Bose-Einstein condensates \cite{eth_2010,eth_jumps,eth_soft}. Additional many-body correlations between the atoms can appear due to collisions or quantum statistics such as Pauli blocking. Collisions can compete with the light forces and give rise to novel Mott and Bose glass phases \cite{maschler_2008,morigi_2010,simons10,hofstetter13,morigi_2013}. Quantum statistics can also significantly alter the self-organization: 
the kinematical constrains imposed by Pauli principle have been shown to modify the scenario more strongly \cite{keeling_2014,piazza_fermi,zhai_2014} than bosonic bunching \cite{piazza_bose}. The effect of Bose condensation on the damping of collective polariton modes in such a setup has also been studied \cite{domokos_damping_one,domokos_damping_two,tureci_2013}.
  

A fundamental open question concerns the nature of thermalization in the non-equilibrium steady state of such systems. 
Take for example the setup sketched in Fig.~\ref{model_system} \cite{vuletic_2003,eth_2010}, wherein 
the balance between coherent drive and cavity decay leads to steady states with non-zero photon number hybridized 
with the atomic gas. Griesser et al.~\cite{griesser_vlasov,niedenzu_2011} and Schuetz et al. \cite{morigi_cooling_2013} have addressed this 
problem (semi-) classically, and argued, based on solutions of classical kinetic equations,
that the atoms attain an effective temperature set by the cavity decay rate $\kappa$.

The purpose of the present paper is to provide the \emph{quantum kinetic theory} for atomic ensembles
in optical resonators and to predict the resulting signatures in the cavity spectrum. 
Our approach, based on the Keldysh path integral \cite{kamenev,dallatorre_2013}, is capable to treat the 
full quantum statistics of (fermionic or bosonic) atoms and cavity decay rates $\kappa$ on equal footing, including situations when 
1/$\kappa$ is the fastest time scale in the problem (bad cavity limit). We now survey our most important results.


\subsection{Key results - atoms}
\label{key_atoms}

In the thermodynamic limit, we find (see Sec.~\ref{sec:atomic_steady_TL}) that, due to the effective infinite-range of the 
photon-mediated atom-atom interactions, the quantum kinetic equation reduces to a Vlasov Equation, independent of quantum statistics:
\begin{align}
\label{QKE_TL_NSC_population_summary} 
\frac{\mathbf{p}}{m}\cdot \mathbf{\nabla}_{\mathbf{X}}n(\mathbf{X},\mathbf{p})-\frac{2\delta_c \lambda^2}{\delta_c^2+\kappa^2}\bigg(\int\frac{d\mathbf{X}^\prime}{V}\sum_{\mathbf{k}}\cos(\mathbf{Q\cdot\mathbf{X}^\prime})n(\mathbf{X}^\prime,\mathbf{k})\bigg)\sin(\mathbf{Q\cdot\mathbf{X}})\mathbf{Q}\cdot \mathbf{\nabla}_{\mathbf{p}}n(\mathbf{X},\mathbf{p})=0\;.
\end{align}
where $n(\mathbf{X},\mathbf{p})$ is the semiclassical steady-state phase-space density of the $N$ atoms in a volume $V$: $\int_V d\mathbf{X}\int d\mathbf{p}n(\mathbf{X},\mathbf{p})=N$, $\lambda$ is the effective strength of the photon-mediated interaction, $\delta_c$ is the (dispersively-shifted) cavity detuning, chosen to be positive for a red-detuned laser, and $\kappa$ the (Markov) decay rate of the cavity mode, the latter being $\cos(\mathbf{Q\cdot x})$. Eq.~(\ref{QKE_TL_NSC_population_summary}), already discussed in \cite{griesser_vlasov} for classical particles in the same setup as the one considered here, is satisfied by any spatially-homogeneous density.
This implies the absence of cavity-driven thermalization of the atomic cloud, which preserves its initial homogeneous phase-space density it had before the coupling to the cavity was turned on. 

Only fluctuations, constituting corrections of order $1/N$ ($\sim$1/volume)  to the Vlasov equation, modify the initial density. This fluctuations become however unstable above a critical coupling strength $\lambda_{sr}$, defining the threshold for the superradiant regime, characterized by a macroscopic spatial modulation of the atomic density together with a finite coherent field in the cavity mode.

\begin{figure}[t]
\includegraphics[width=100mm]{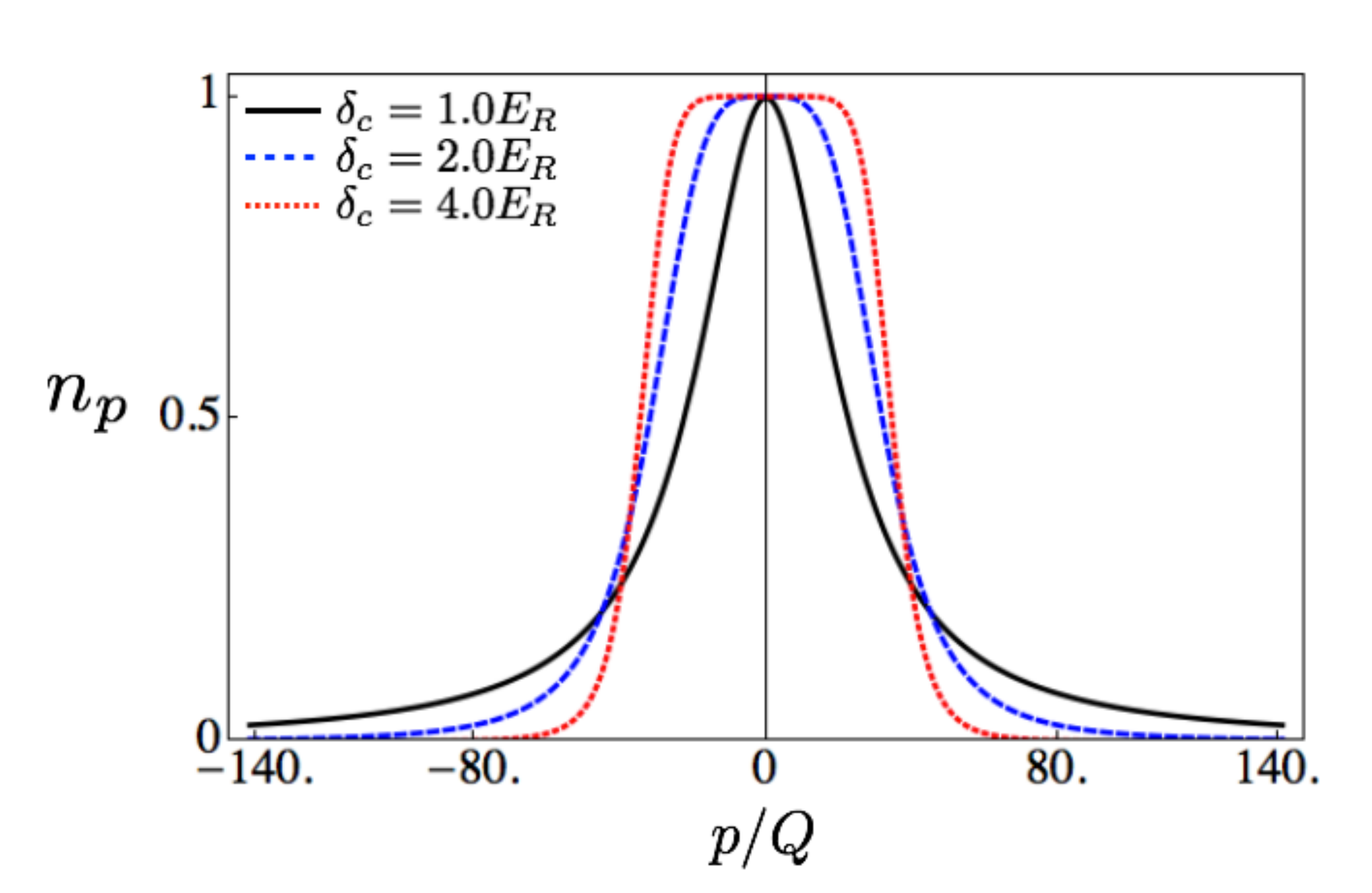}
\caption{Steady state distribution for a fermionic gas coupled to a cavity mode decaying with $\kappa=2\; E_R$. For increasing values of the detuning, the distribution tends to a thermal Fermi-Dirac distribution with the effective temperature Eq.~(\ref{Teff_atoms_summary}).}
\label{QKE_dist}
\end{figure}

Outside the superradiant phase, the quantum kinetic equation including the fluctuations up to order $1/N$ contains a collisional term leading to cavity-driven thermalization of the atoms (see Sec.~\ref{sec:one_over_N_QKE}), governed by the following equation for the spatially-averaged atomic phase-space density $n^{(0)}_{\mathbf{p}}$:
\[
\left(\delta_c^2+\kappa^2+\omega_{\mathbf{Q}}(\mathbf{p})\right)(n^{(0)}_{\mathbf{p+Q}}-n^{(0)}_{\mathbf{p}})=-2\delta_c \omega_{\mathbf{Q}}(\mathbf{p})\left[n^{(0)}_{\mathbf{p+Q}}+n^{(0)}_{\mathbf{p}}\pm 2 n^{(0)}_{\mathbf{p+Q}}n^{(0)}_{\mathbf{p}}\right]
\]
where the $+(-)$ refers to bosons(fermions) and with the particle-hole dispersion $\omega_{\mathbf{Q}}(\mathbf{p})=(Q^2/2m+\mathbf{Q\cdot p}/m)$
. For a smooth density on the recoil scale $E_R=Q^2/2m$, that is $n^{(0)}_{\mathbf{p+Q}}\simeq n^{(0)}_{\mathbf{p}}+\mathbf{Q}\cdot\boldsymbol{\nabla_p}n^{(0)}_{\mathbf{p}}$ with $\mathbf{Q}\cdot\boldsymbol{\nabla_p}n^{(0)}_{\mathbf{p}}\ll n^{(0)}_{\mathbf{p}}$, the above equation has a unique non-thermal steady-state solution containing the effects of quantum statistics
\begin{align}
\label{gen_tsallis_summary}
n_p^{(0)}=\frac{1}{C\left(1+4\frac{E_R\epsilon_p}{\delta_c^2+\kappa^2}\right)^{\frac{\delta_c}{E_R}}\mp 1}\;,
\end{align}
with $C$ a normalization constant and $\epsilon_p=p^2/2m$. 
The distribution power-law (\ref{gen_tsallis_summary}) is depicted in Fig.~\ref{QKE_dist} for a Fermi gas. 
In the limit $\delta_c\gg E_R$, the distribution (\ref{gen_tsallis_summary}) tends to a thermal Bose(Fermi) momentum distribution, with an effective temperature set by
\begin{equation}
\label{Teff_atoms_summary}
k_B T_{\rm eff}^{(\rm at)}=\frac{\delta_c^2+\kappa^2}{4\delta_c}\;.
\end{equation}
 In the limit of small densitites $n_p^{(0)}\ll 1$ where the quantum-statistical effects disappear, this distribution tends to the Tsallis distribution, in accordance with the results obtained for a classical gas \cite{griesser_vlasov,niedenzu_2011,morigi_cooling_2013}.
The effective temperature \eqref{Teff_atoms_summary} coincides with the effective temperature of the $X$-component of a harmonic oscillator of frequency $\delta_c$ coupled to a Markov bath at a rate $\kappa$ after tracing out the $P$-component.
The atoms couple indeed only to the $X$ component of the cavity and the above temperature can be transferred by the nonlinearities resulting from the fluctuations of order $1/N$. This also happens when the atoms are subjected to quenched disorder \cite{buchhold_2013}.

The systematic expansion in $1/N$ for the steady state relies on a particular choice of the order in which the thermodynamic limit $N,V\to\infty$ and the long time limit $t\to\infty$ are taken, namely the former before the latter. However, in any realistic (and therefore finite) system 
this will not be the relevant order in which to take the limits. Most importantly, these limits do not in general commute, since the absence of an
equilibrium fluctuation-dissipation theorem forces us to determine the long time steady state distribution in a self-consistent way. 
Indeed, even though scaling like $1/N$, the fluctuations will eventually become relevant at long enough times. 
Perturbative control can be achieved sufficiently far away from threshold ($\lambda$ relatively small),
at large $N$, and restricting to moderately short times. Starting from such a state, the thermalization rate of the atoms resulting from the quantum kinetic equation at order $1/N$ (see Sec.~\ref{sec:one_over_N_QKE}):
\begin{align}
\label{thermalization_rate}
\Gamma_{\rm th}^{(1/N)}\simeq \frac{\lambda^2\delta_c\kappa E_R}{(\delta_c^2+\kappa^2)^2}\xrightarrow{\kappa\gg\delta_c}\frac{\lambda^2\delta_c E_R}{\kappa^3}\;,
\end{align}
where $\lambda^2\propto 1/N$ in the thermodynamic limit, so that $\Gamma_{\rm th}^{(1/N)}$ scales like $1/N$. Therefore, at times of the order of $t^{(1/N)}\simeq \kappa^3/\lambda^2\delta_c E_R\propto N$ the fluctuations leading to atomic thermalization become important and our quantum kinetic equation predicts the atoms to attain the distribution (\ref{gen_tsallis_summary}).

However, the quantum kinetic equation leading to (\ref{gen_tsallis_summary}) breaks down close to the superradiant threshold,
where the collective excitations become soft and the resulting slowing down of the dynamics requires a self-consistent determination of the photon steady-state distribution together with the atomic one. This additional effect is not included in the derivation of the atomic distribution \eqref{gen_tsallis_summary}, which would be therefore valid only away from threshold, where the photon dynamics is weakly hybridized with the atoms.
This is justified as long as the photon's scattering rate with the medium $\Gamma_{\rm scatter}$ is smaller 
than the photon decay rate $\kappa$. In our set-up, this means
\begin{align}
\kappa \gg \Gamma_{\rm scatter} \approx\frac{ N\lambda^2}{ E_R}\leftrightarrow \epsilon\equiv\frac{N\lambda^2}{\kappa E_R}\ll1
\label{eq:condition}
\end{align}
where we took the recoil energy $E_R$ to be the typical atomic energy scale (in general we have to calculate the polarization function of the medium, see Sec.~\ref{subsec:photons}).
With $\kappa\sim$ MHz, $\lambda = g_0 \Omega/ \Delta_a \sim$ kHz and $E_R\sim$ kHz, this typically holds
sufficiently far away from superradiance threshold. The latter is indeed approximately determined as $\lambda_{sr}^2 \approx \frac{\kappa E_R}{N}$, so that the condition (\ref{eq:condition}) is never fulfilled close to threshold where $\epsilon\simeq 1$. 
We point out that the conditions justifying the non-self-consistent determination of the atomic together with the photonic distribution depend on the particular setup and aren't necessarily connected to the presence of a threshold. In dye-filled optical microcavities for instance \cite{weitz_photoncond_2010,keeling_photoncond_2013,stoof_photoncond_2013,sela_2014}, the dye molecule decoherence rate is by far the fastest scale and can induce a strong thermalization of the photons with negligible back-action on the molecular distribution.

In the future, we note that the full self-consistent determination of the steady-state photon distribution required close to threshold would be possible within our Keldysh approach (see Appendix \ref{app:self}) by including the polarization correction to the photon Green functions (both retarded and Keldysh), renormalizing the cavity-mediated atom-atom interaction with a term which functionally depends on the atomic distribution itself. This, the situation in which the photon recoil momentum $\mathbf{Q}$ is large (and the slow approximation 
Eq.~(\ref{eq:slow}) breaks down), and an explicit forward integration in time 
of the coupled quantum kinetic equations are interesting topics for future papers. 


\subsection{Key results - photons}
\label{key_photons}

In the light of what we discussed in the previous section, we hold time fixed and assume to have a system large enough to be able to neglect the $1/N$ cavity-driven thermalization of the atoms. The atoms are in a thermal Bose or Fermi momentum distribution with a given (independent) temperature $T$, achieved by standard cooling methods for the ultracold gas. 

In this case, we calculate several properties of the steady state of the cavity field, driven by pump photons scattered from the atomic cloud and decaying into the Markov vacuum bath out of the cavity mirrors.

\begin{figure}[t]
\includegraphics[width=160mm]{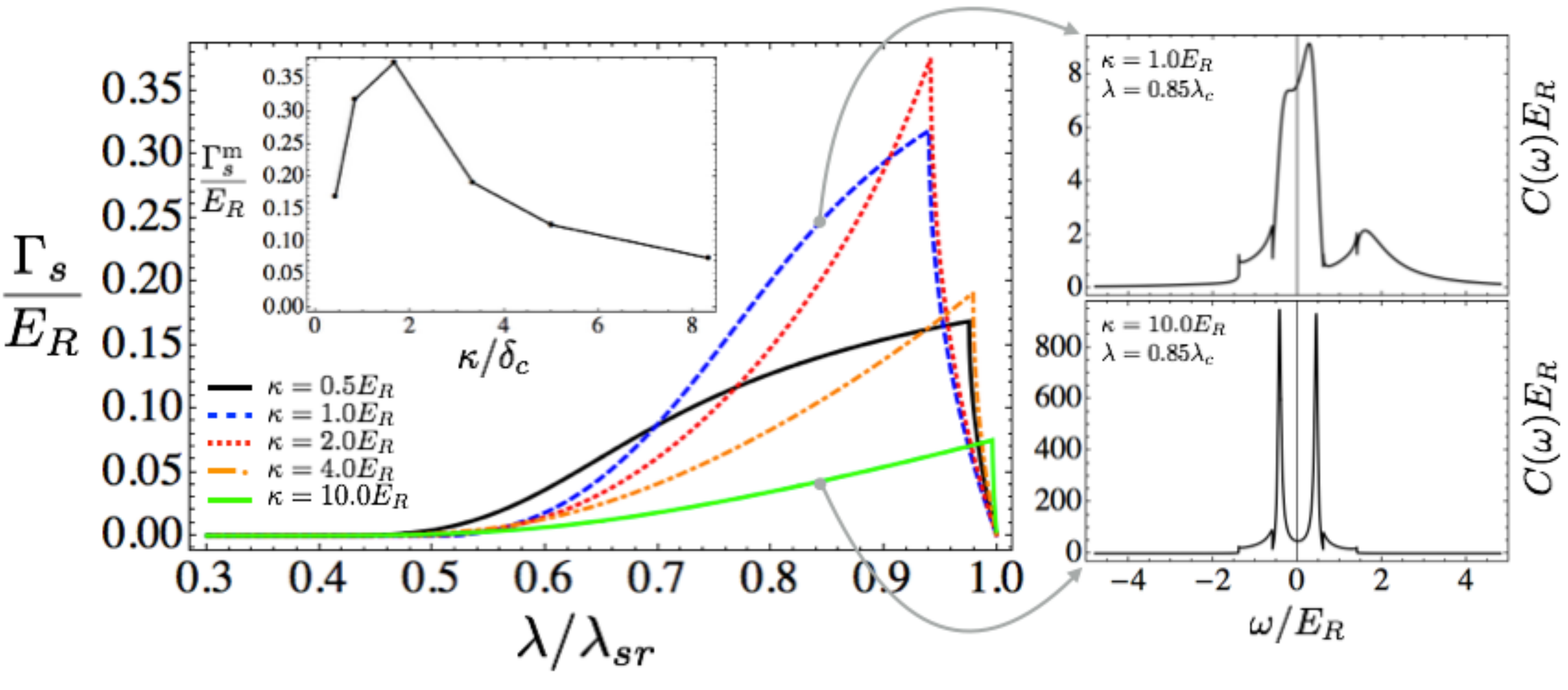}
\caption{Left: damping rate (imaginary part of the eigenmode) of the lower polariton collective mode as a function of the two-photon atom-cavity coupling, for increasing values of the cavity decay rate. Inset: highest damping rate as a function of the ratio of cavity decay to detuning. The damping rate increases with increasing coupling, up to a maximum value (coinciding with the vanishing of the real part of the eigenmode) after which it decreased to finally vanish at the superradiant threshold. Initially, the role of cavity decay is to increase the polariton damping as long as it is of the order of the cavity detuning. Beyond this value we reach the bad-cavity regime where the polariton damping rate decreases with cavity decay. This decrease is restricted to values of the coupling up to the point where the soft-mode becomes purely dissipative, marked by the maximum in the damping as a function of the coupling.
There is no additional polariton damping source apart from cavity decay since for the degenerate Fermi gas Landau damping is suppressed for frequencies outside the particle-hole continuum. Right: two examples of the intracavity photon spectral density inside and outside the bad-cavity regime. 
Here we chose an almost resonant case $\delta_c=1.2\;E_R$. Away from resonance $\delta_c\gg\;E_R$ the polariton damping is even less affected by cavity decay but the qualitative behavior above remains. We chose a low density regime $k_F=0.2Q$. }
\label{damping}
\end{figure}
\begin{figure}[t]
\includegraphics[width=160mm]{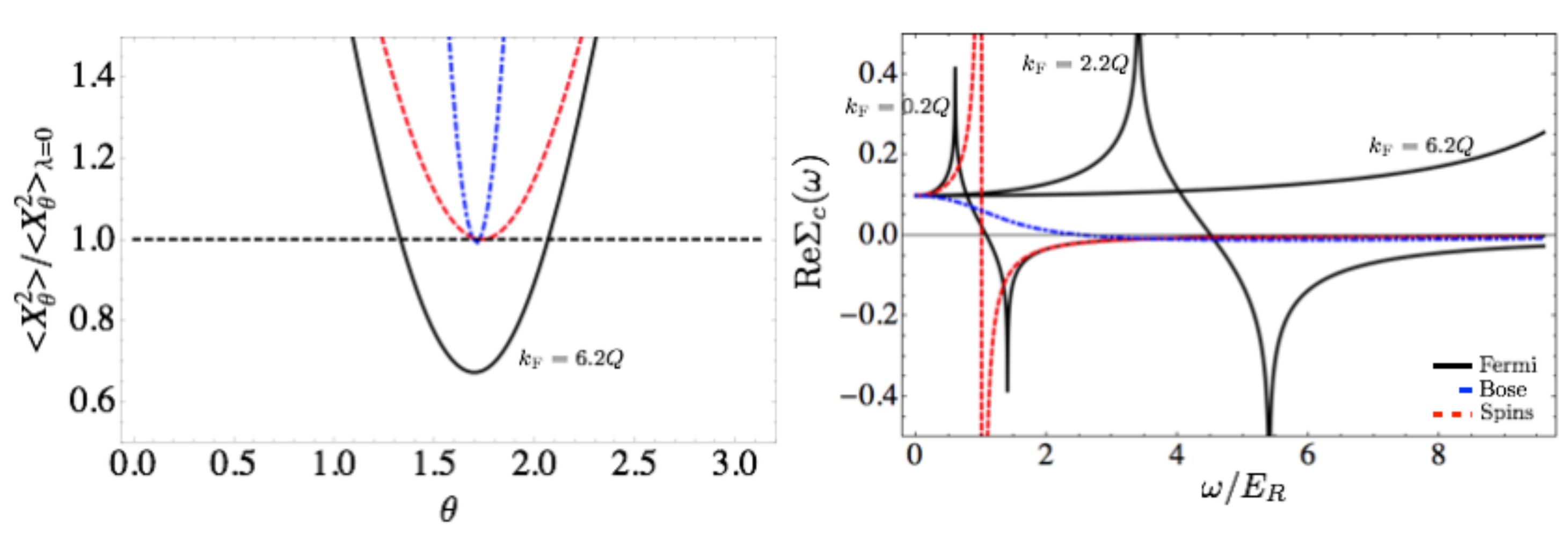}
\caption{
Left: comparison of  the equal time quadrature variance at coupling $\lambda=0.9\lambda_{\rm sr}$ for three different systems: $1d$ fermi gas at $T=0$ (black solid line), spins \cite{dallatorre_2013} (red dashed line), $3d$ bose gas $T=1.1T_{\rm bec}$ (blue dash-dotted line). Parameters are $\kappa=0.2 E_{\rm R}$ and $\delta_{\rm c}=1.2 E_{\rm R}$ with a large density of the $1d$ fermi gas: $k_{\rm F}=6.2 Q$. The Fermi gas induces squeezing of the quadrature variance below the vacuum shot-noise (black-dashed line). Right: corresponding behavior of the photonic dispersion in the atomic medium, which corresponds to the effective frequency-dependent driving strength of the cavity-mode, achieved by coherently driving the atoms. Squeezing is achieved at high fermionic density because the frequency dependence of the dispersion in the atomic medium is suppressed due to Pauli principle. This induces an effective flat driving strength, analog to the one employed to achieve squeezing in an optical parametric oscillator \cite{carmichael_book}.}
\label{quadrature_large_density}
\end{figure}

First we calculate the frequency distribution function $f_{ph}(\omega)$ of the photons (of the full field, not the $X$-component) scattered into the cavity mode and compare fermions with non-interacting non-condensed bosons and spins (see Sec.~\ref{sec:photon_LET}). We find in general thermal behavior $f(\omega)\simeq 2T_{\rm eff}^{(\rm ph)}/\omega$ at low frequency, with an effective temperature $T_{\rm eff}$ which depends nontrivially on the dispersion/absorption properties of the atomic medium. For instance, taking spins \cite{dallatorre_2013} and low-$T$ fermions away from nesting $Q=2k_F$ as the medium, both showing no absorption at low frequencies, we find the effective temperature
\[
T_{\rm eff}^{(\rm ph)}=\frac{1}{4}\lambda^2\mathrm{Re}\Pi^R(0,Q)\;,\;\;\text{non-nested fermions or spins}\;,
\]
with the real(imaginary) part of the polarization function $\Pi^R(\omega,k)$ characterizing the medium dispersion(absorption). For bosons, showing an absorption linearly vanishing with frequency: $\mathrm{Im}\Pi^R(\omega,k)\propto \omega$, the effective temperature reads instead
\[
T_{\rm eff}^{(\rm ph)}=\frac{1}{4}\lambda^2\frac{\mathrm{Re}\Pi^R(0,Q)}{\sqrt{1+\lambda^4 \mathrm{Im}^2\Pi^R(\omega,Q)/\omega^2}} \;,\;\;\text{bosons}\;.
\]
We note that $T_{\rm eff}^{(\rm ph)}$ depends on the initial atomic temperature $T$ in a complicated way through the polarization function.
On the other hand, the case of a one-dimensional Fermi cloud at perfect nesting $Q=2k_F$ is exceptional in this respect due to a frequency-independent absorption down to zero frequency, which induces a non-thermal behavior of the photon field:
$f_{ph}^{Q=2k_{\rm F}}(\omega)\xrightarrow{\kappa\gg E_{\rm R},\omega\to 0}\frac{\sqrt{\mathrm{Re}\Pi^{R^2}(0,Q)+\mathrm{Im}\Pi^{R^2}(0,Q)}}{\mathrm{Im}\Pi^{R}(0,Q)}=\text{const.}$.

We also discuss the properties of the cavity spectrum (see Sec.~\ref{sec:spectral}) and show in particular that in the bad-cavity limit $\kappa \gg E_R$ the linewidth of the polaritonic sidebands is limited only by the atomic absorption, the latter being exponentially small for a low-temperature collisionless Fermi gas away from perfect nesting, due to the reduced phase-space for Landau damping, as observable in the narrowed peaks 
on the right of Fig.~\ref{damping}. The absence of atomic absorption in certain frequency windows is peculiar to the collisionless Fermi gas. For bosonic gases, where collisions are important at low temperature, the damping due to the atomic medium is not exponentially suppressed. This has been computed with other methods \cite{domokos_damping_one,domokos_damping_two,tureci_2013}and also measured experimentally \cite{eth_non_eq}.
%

We finally calculate the equal-time quadrature variance of the cavity light in Fig.~\ref{quadrature_large_density}. Differently from a Bose gas or spins, the degenerate Fermi gas can generate up to $50\%$ intra-cavity squeezing close to the superradiant threshold, due to the Pauli principle causing the suppression of the frequency dependence of the scattered photons for large atomic densities.

\section{Coupled atom-photon model}

In this section, we present the model for the driven-dissipative 
atom-photon system Fig.~\ref{model_system}. First, as a periodically driven 
Hamiltonian supplemented with Lindblad terms for cavity decay. Then the 
equivalent Keldysh action. 

\subsection{Hamiltonian}
Due to the periodic driving of the pump laser with $\sim\Omega e^{-i \omega_p t}$
the Hamiltonian for Fig.~\ref{model_system} is inherently time-dependent $H(t)$.
However, it is convenient to go to a frame rotating with $\omega_p$ which is 
a (fast) optical frequency \cite{carmichael_2007,maschler_2008}. In this frame,
the explicit time-dependence is gone, however this turns the bath of 
cavity modes into a Markovian bath \cite{dallatorre_2013}.
In terms of the quantized field
operators $\hat{\psi}_{g/e}$ for the atoms in the internal ground or excited state
and the annihilation operator $\hat{a}$ for a cavity photon,
the complete atom plus driven cavity Hamiltonian reads~\cite{maschler_2008}
\begin{equation}\label{hamiltonian}
\hat{H}=\hat{H}_{\rm A}+\hat{H}_{\rm C}+\hat{H}_{\rm AC}+\hat{H}_{\rm
  AP}\ , 
\end{equation}
where
\begin{align}
\hat{H}_{\rm A}&=\int
d\mathbf{r}\left[\hat{\psi}_g^{\dag}(\mathbf{r})(-\frac{\nabla^2}{2
    m})\hat{\psi}_g^{}(\mathbf{r})+\hat{\psi}_e^{\dag}(\mathbf{r})(-\frac{\nabla^2}{2
    m}-\Delta_{\rm a})\hat{\psi}_e^{}(\mathbf{r})\right]\nonumber\\
\hat{H}_{\rm C}&=-\Delta_{\rm c}\ \hat{a}^{\dag}\hat{a}^{}\nonumber\\
\hat{H}_{\rm AC}&=-i\ g_0\int d\mathbf{r} \hat{\psi}_g^{\dag}(\mathbf{r})\eta_{\rm c}(\mathbf{r})\hat{a}^{\dag}\hat{\psi}_e^{}(\mathbf{r})+ {\rm h.c}\nonumber\\
\hat{H}_{\rm AP}&=-i\ \Omega\int d\mathbf{r} \hat{\psi}_g^{\dag}(\mathbf{r})\eta_{\rm p}(\mathbf{r})\hat{\psi}_e^{}(\mathbf{r})+ {\rm h.c}\nonumber
\end{align}
in the frame rotating with the pump frequency
$\omega_{\rm p}$.  Here, $\Delta_{\rm a}=\omega_{\rm p}-\omega_{\rm e}$ and
$\Delta_{\rm c}=\omega_{\rm p}-\omega_{\rm c}$ are the detunings between the pump and
the atomic resonance, and the pump and the cavity mode,
respectively (we set $\hbar=1$ except in some final results). Moreover, $m$ is the atomic mass, $g_0$ is the single-photon Rabi coupling between
the atom and the cavity and $\Omega$ is the pump Rabi frequency. 
The functions $\eta_{\rm c}(\mathbf r), \eta_{\rm p}(\mathbf r)$ contain the
spatial form of the cavity and pump modes, respectively.
In the following, we consider the large detuning regime, where $1/\Delta_{\rm a}$ is the fastest time-scale. This allows us to neglect spontaneous emission from the excited atomic level and also to adiabatically eliminate the latter, to obtain the following effective Hamiltonians:
\begin{align}
\hat{H}_{\rm eff,A}&=\int d\mathbf{r}\,
\hat{\psi}^{\dag}(\mathbf{r})
\left\{
-\frac{\nabla^2}{2m}
+
\frac{\Omega^2\eta_{\rm p}^2(\mathbf{r})}{\Delta_a}
\right\}
\hat{\psi}^{}(\mathbf{r})
\nonumber\\
\hat{H}_{\rm eff,C} &= -\Delta_{\rm c}\ \hat{a}^{\dag}\hat{a}^{}
\nonumber\\
\hat{H}_{\rm eff,AC} &= 
\int d\mathbf{r}\,
\hat{\psi}^{\dag}(\mathbf{r})
\Bigg\{\frac{\left(g_0 \eta_{\rm c}(\mathbf{r})\right)^2}{\Delta_a}\hat{a}^\dagger\hat{a}+\frac{ \Omega g_0\eta_{\rm c}(\mathbf{r})\eta_{\rm p}(\mathbf{r})}{\Delta_a} 
\left(\hat{a} + \hat{a}^\dagger \right)
\Bigg\} 
\hat{\psi}^{}(\mathbf{r})\;,
\label{eq:H_eff}
\end{align}
where we suppressed the subscript $g$. 
 Atomic spontaneous emission is neglected due to the large detuning. Cavity decay is included within the usual Markov approximation, leading to the Lindblad term in the master equation \cite{maschler_2008}
\begin{equation}
\mathcal{L}\hat{\rho}=\kappa\left(2\hat{a}\hat{\rho}\hat{a}^\dag-\hat{a}^\dag\hat{a}\hat{\rho}-\hat{\rho}\hat{a}^\dag\hat{a}\right)\;.
\label{lindblad}
\end{equation}
%

\subsection{Keldysh action}
In order to properly take into account the non-unitary dynamics introduced by the transversal drive and 
cavity decay \eqref{lindblad}, we now formulate the model introduced in the preceding section as an action on the 
Keldysh closed time-contour $\mathcal{C}$ \cite{dallatorre_2013,kamenev}. 
In this section, we specify the case of fermionic atoms (the construction for bosonic atoms proceeds 
analogously).

Correlators of the atoms and photons can be obtained from the generating functional
\begin{equation}
Z=\frac{1}{\mathrm{Tr}[\hat{\rho}_0]}\int \mathcal{D}\{\bar{\psi}\}\mathcal{D}\{\psi\}\mathcal{D}a^*\mathcal{D}a \;e^{iS\left[\bar{\psi},\psi,a^*,a\right]}\;,
\end{equation}
with the action $S\left[\bar{\psi},\psi,a^*,a\right]=S_0+S_V$,
\begin{align}
\label{action_full}
S_0\left[\bar{\psi},\psi,a^*,a\right]=&\oint_{\mathcal{C}}dt \int d\mathbf{r}\left[\bar{\psi}(\mathbf{r},t)i\partial_t\psi(\mathbf{r},t)-
H_{\rm A}(\bar{\psi},\psi)\right]
+
\oint_{\mathcal{C}} dt \left[a^*(t)i\partial_ta(t)-H_{\rm C}(a^*,a)\right]\;,\nonumber\\
S_V\left[\bar{\psi},\psi,a^*,a\right]=&-\oint_{\mathcal{C}}dt \int d\mathbf{r}\;\bar{\psi}(\mathbf{r},t)\psi(\mathbf{r},t)V_{a^\ast a}(\mathbf{r},t)\;,
\end{align}
where the photonic part of the interaction term is
\begin{align}
V_{a^\ast a}(\mathbf{r},t)=&\frac{\left(g_0 \eta_{\rm c}(\mathbf{r})\right)^2}{\Delta_a}a^*(t)a(t) 
+
\frac{\Omega g_0\eta_{\rm c}(\mathbf{r})\eta_{\rm p}(\mathbf{r})}{\Delta_a}\left(a(t) + a^*(t)\right)\;.
\end{align}
Here $a,a^*(\psi,\bar{\psi})$ denote complex (Grassmann) fields, and the initial density matrix $\hat{\rho}_0=\hat{\rho}_{\rm 0,A}\otimes \hat{\rho}_{{\rm 0,C}}=\exp(-\beta\hat{H}_{\rm eff,A})\otimes\exp(-\beta\hat{H}_{\rm eff,C})$ corresponds to the uncoupled system, so that $Z|_{V=0}=1$.

Following the usual procedure, we split $\mathcal{C}$ into forward $+$ and backward $-$ contours and subsequently perform the \emph{fermionic} Keldysh rotation 
for the atomic field and the \emph{bosonic} one for the cavity field:
\begin{align}
\psi_1=\frac{1}{\sqrt{2}}(\psi_{+}+\psi_{-})\;,\;\;\psi_2=\frac{1}{\sqrt{2}}(\psi_{+}-\psi_{-})\;,\;\;
\bar{\psi}_1=\frac{1}{\sqrt{2}}(\bar{\psi}_{+}-\bar{\psi}_{-})\;,\;\;\bar{\psi}_2=\frac{1}{\sqrt{2}}(\bar{\psi}_{+}+\bar{\psi}_{-})\nonumber\\
a_{cl}=\frac{1}{\sqrt{2}}(a_{+}+a_{-})\;,\;\;a_{q}=\frac{1}{\sqrt{2}}(a_{+}-a_{-})\;,\;\;a_{cl}^*=\frac{1}{\sqrt{2}}(a_{+}^*+a_{-}^*)\;,\;\;a_{q}^*=\frac{1}{\sqrt{2}}(a_{+}^*-a_{-}^*)\;.
\end{align}
In this basis, by performing space and time Fourier transforms, we can rewrite the action
\begin{align}
\label{eq:bare_action}
&S_0[\bar{\psi}_{1,2},\psi_{1,2},a_{cl,q}^*,a_{cl,q}]=S_{\rm 0,C}[a_{cl,q}^*,a_{cl,q}]+S_{\rm 0,A}[\bar{\psi}_{1,2},\psi_{1,2}]
=\\
&\int_{\infty}^\infty\frac{d\omega}{2\pi}\mathbf{a}^\dag(\omega)\cdot\left(\begin{array}{cc} 0 &
 \omega+\Delta_{\rm c}-i\kappa\\ \omega+\Delta_{\rm c}+i\kappa &2i\kappa\end{array}\right)\cdot\mathbf{a}(\omega)
+
\int_{\infty}^\infty\frac{d\omega}{2\pi}\sum_{\substack{\mathbf{k}}}\bar{\mathbf{\Psi}}^T\!(\omega,\mathbf{k})\cdot\underline{\mathbf{G}}_0^{-1}(\omega,\mathbf{k})\cdot\mathbf{\Psi}(\omega,\mathbf{k})\;,\nonumber
\end{align}
with $\Delta_c < 0$, the vectors $\mathbf{a}(\omega)^T=\left(a_{cl}(\omega),a_q(\omega)\right)$, $\mathbf{\Psi}^T(\omega,\mathbf{k})=\left(\psi_1(\omega,\mathbf{k}),\psi_2(\omega,\mathbf{k})\right)$, and the inverse free atom propagator 
\begin{equation}
\underline{\mathbf{G}}_0^{-1}(\omega,\mathbf{k})=\left(\begin{array}{cc} [G_0^R(\omega,\mathbf{k})]^{-1}& [G_0^{-1}(\omega,\mathbf{k})]^{K}\\0 & [G_0^A(\omega,\mathbf{k})]^{-1}\end{array}\right)\;
\end{equation}
where
\begin{align}
\label{atom_rak_explicit}
G_0^{R(A)}(\omega,\mathbf{k})=\frac{1}{\omega-\epsilon_{\mathbf{k}}\pm i0^+}\;,\;\; G_0^K (\omega,\mathbf{k})=-2\pi i\;F(\omega)\delta(\omega-\epsilon_{\mathbf{k}})\;,
\end{align}
with the free atomic dispersion $\epsilon_{\mathbf{k}}$ \emph{without} the chemical potential $\mu$. When the atoms are in equilibrium at temperature $T$ we have 
\begin{equation}
\label{dist_eq}
  F_0^{\rm (eq)}(\omega)=1-2n_{\rm F}(\omega)=\tanh\left(\frac{\omega-\mu}{2T}\right)\;.
\end{equation}
Since the interaction part $S_V$ involves $a(t)+a^*(t)$, it is convenient to rewrite the photon propagator from Eq.~(\ref{eq:bare_action}) in the vector notation
\begin{equation}
S_{\rm 0}[a_{cl,q}^*,a_{cl,q}]=\frac12\int_{-\infty}^\infty\frac{d\omega}{2\pi}\left(\begin{array}{cccc}a_{cl}^*(\omega) & a_{cl}(-\omega) & a_{q}^*(\omega) & a_{q}(-\omega)\end{array}\right)\cdot\left(\begin{array}{cc}0 & g_{0,2x2}^{A^{-1}}(\omega)\\ g_{0,2x2}^{R^{-1}}(\omega) & d_{0,2x2}^K(\omega)\end{array}\right) \cdot\left(\begin{array}{c}a_{cl}(\omega) \\ a_{cl}^*(-\omega) \\ a_{q}(\omega) \\ a_{q}^*(-\omega)\end{array}\right)\;,
\end{equation}
with $\delta_c = - \Delta_ c + 1/2 U_0 N$, 
\begin{equation}
\label{GR_cavity}
g_{0,2x2}^{R^{-1}}(\omega)
=
\left(\begin{array}{cc}\omega-\delta_c+i\kappa 
& 0 \\ 0 & 
-\omega-\delta_c-i\kappa \end{array}\right)\;, 
\end{equation}
and $g_{0,2x2}^{A^{-1}}(\omega)=[g_{0,2x2}^{R^{-1}}(\omega)]^\dag$. 
The bare Keldysh component of the photons reads 
\begin{equation}
d_{0,2x2}^K(\omega)\left(\begin{array}{cc}2i\kappa  & 0 \\ 0 & 2i\kappa \end{array}\right)\;.
\end{equation}
The interaction part can also be written as a $2x2$ matrix
\begin{align}
&S_V[\bar{\psi}_{1,2},\psi_{1,2},a_{cl,q}^*,a_{cl,q}]=
-\int_{\infty}^\infty\frac{d\omega d\omega^\prime}{(2\pi)^2}
\sum_{\substack{\mathbf{k},\mathbf{k}^\prime}}\bar{\mathbf{\Psi}}^T\!(\omega,\mathbf{k})\cdot\underline{\mathbf{V}}(\omega-\omega^\prime,\mathbf{k}-\mathbf{k}^\prime)\cdot\mathbf{\Psi}(\omega^\prime,\mathbf{k}^\prime)\;,
\end{align}
where
\begin{equation}
\underline{\mathbf{V}}(\omega,\mathbf{k})=\left(\begin{array}{cc} V_{cl}(\omega,\mathbf{k})& V_q (\omega,\mathbf{k})\\V_q (\omega,\mathbf{k}) &V_{cl}(\omega,\mathbf{k})\end{array}\right)\;,
\end{equation}
with $V_{cl(q)}=(V_+\pm V_-)/2$, so that, explicitly, we have for the classical component 
\begin{align}
\label{sources_explicit}
V_{cl}(\omega,\mathbf{k})=&\frac{1}{2}\frac{g_0^2}{\Delta_{\rm a}}\eta_{\rm CC}(\mathbf{k}) 
\int_{\omega'}
\Bigg(a_{cl}^*(\omega^\prime)a_{cl}(\omega^\prime +\omega)+a_{q}^*(\omega^\prime)a_{q}(\omega^\prime+\omega)\Bigg)+\frac{1}{\sqrt{2}}\frac{g_0\Omega}{\Delta_{\rm a}}\eta_{\rm PC}(\mathbf{k})\left(a_{cl}^*(-\omega)+a_{cl}(\omega)\right)\;,
\end{align}
and for the quantum component
\begin{align}
V_{q}(\omega,\mathbf{k})=\frac{1}{2}\frac{g_0^2}{\Delta_{\rm a}}\eta_{\rm CC}(\mathbf{k})
 \int_{\omega'}
 \left(a_{cl}^*(\omega^\prime)a_{q}(\omega^\prime+\omega)+a_{q}^*(\omega^\prime)a_{cl}(\omega^\prime+\omega)\right)+\frac{1}{\sqrt{2}}\frac{g_0\Omega}{\Delta_{\rm a}}\eta_{\rm PC}(\mathbf{k})\left(a_{q}^*(-\omega)+a_{q}(\omega)\right)\;,
\end{align}
where the geometric factors $\eta_{\rm CC}(\mathbf{k})=\int d\mathbf{r}\exp(i\mathbf{k}\cdot\mathbf{r})\eta_{\rm c}^2(\mathbf{r})$, $\eta_{\rm PC}(\mathbf{k})=\int d\mathbf{r}\exp(i\mathbf{k}\cdot\mathbf{r})\eta_{\rm p}(\mathbf{r}) \eta_{\rm c}(\mathbf{r})$ describe the scattering from the cavity into the cavity and from the pump into the cavity, respectively. 


\section{Quantum kinetics of atoms}

In this section, we derive the quantum kinetic equation for the atoms. To achieve this, 
we begin by deriving the Dyson equation and the corresponding self-energy diagrams for the atoms 
in a general form. Wherever possible, we follow Kamenev's notation \cite{kamenev}.
We then set up the $1/N$ expansion where $N$ is the number of atoms such that the leading order 
corresponds to the thermodynamic limit (TL) $N\rightarrow \infty,V\rightarrow \infty$ at $N/V=\text{const.}$. We finally present solutions for the 
distribution functions of the atoms first at $N\rightarrow \infty$ and then at next-to-leading order in 
$1/N$.
The interaction energy is finte in TL because the dipole coupling constant $g_0$ is proportional to $V^{-1/2}$ and therefore the couplings scale as
\[
\lambda\equiv\frac{g_0\Omega}{\Delta_a}\propto V^{-1/2}\;,\;\;\;U_0\equiv\frac{g_0^2}{\Delta_a}\propto V^{-1}\;.
\]

\subsection{Dyson equation}

The starting point is the Dyson equation for the atom matrix propagator $\underline{G}$:
\begin{equation}
\label{dyson_matrix}
\left(\underline{G}_0^{-1}-\underline{\Sigma}\right)\circ\underline{G}=\underline{1}\;,
\end{equation}
with the free atom propagator $\underline{G}_0$ and the self-energy  $\underline{\Sigma}$ resulting from the cavity-mediated atom-atom interactions. The latter \emph{always} possesses the causality structure
\[
\underline{\Sigma}=\left(\begin{array}{cc} \Sigma^R & \Sigma^K \\ 0 & \Sigma^A \end{array}\right)\;.
\]
It has to be determined from the interaction vertex $S_V$, given in Eq.~(\ref{action_full}), in some approximation. The essential observation is that $\Sigma^{(R,A,K)}$ in general depend on the atomic distribution $F$, as we will see later. The Dyson equation can be thus rewritten component by component as

\begin{align}
\left(i\partial_t+\frac{\nabla^2}{2m}-\Sigma^{R(A)}\circ\right)G^{R(A)}(x-x^\prime)&=\delta(x-x^\prime)\;,\label{dyson_RA}\\
F\circ G^{A^{-1}}(x,x^\prime)-G^{R^{-1}}\circ F (x,x^\prime)&=\Sigma^K(x,x^\prime)-\left(\Sigma^{R}\circ F (x,x^\prime) -F\circ \Sigma^{A}(x,x^\prime)\right)\;,\label{BE}
\end{align}
with the 4-coordinate $x=(x_t,\mathbf{x})$, where the symbol $\circ$ stands for spacetime convolution, and with the usual parametrization 
\begin{equation}
\label{GK_par}
G^K(x,x^\prime)=G^{R}\circ F (x,x^\prime) -F\circ G^{A}(x,x^\prime)\;.
\end{equation}

A convenient strategy to obtain the self-energy contractions that enter the Dyson equation
Eq.~(\ref{BE}) is a cumulant expansion of the interaction vertex 
\[
\Phi=\langle e^{i S_V}\rangle_0\;,
\]
where the average is performed with respect to the non-interacting part of the action. 
The interacting part of the action $S_V$, given in Eq.~(\ref{action_full}), can be conveniently collected 
in a matrix notation 
\[
S_V=-\int dx\sum_{a,b=1,2}\sum_{\alpha=cl,q}\bar{\psi}_a(x)V_\alpha(x)\gamma_{ab}^\alpha\psi_b(x)\;,
\]
with the 2-by-2 matrices
\[
\underline{\gamma}^{cl}=\left(\begin{array}{cc} 1 & 0 \\ 0 & 1 \end{array}\right)\;,\;\;\underline{\gamma}^q=\left(\begin{array}{cc} 0 & 1 \\ 1 & 0 \end{array}\right)\;.
\]
The self-energy is then given by
\begin{equation}
\Sigma_{ab}(x,x^\prime)=\frac{\delta\Phi}{\delta G_{ba}(x^\prime,x)}\;,
\quad
G_{ab}(x,x^\prime)=-i\langle\psi_a(x)\bar{\psi}_b(x^\prime)\rangle\;,
\end{equation}
and the causality structure $G_{11}=G^R$, $G_{22}=G^A$, $G_{12}=G^K$, and $G_{21}=0$ (the same for $\Sigma$).

In the following, we restrict to the non-superradiant phase where the macroscopic mean-field contribution to the cavity field is absent.
We expand $\Phi$ in powers of $V$ and truncate up to second order in $a_{cl,q}$. 

\begin{figure}[t]
\includegraphics[width=150mm]{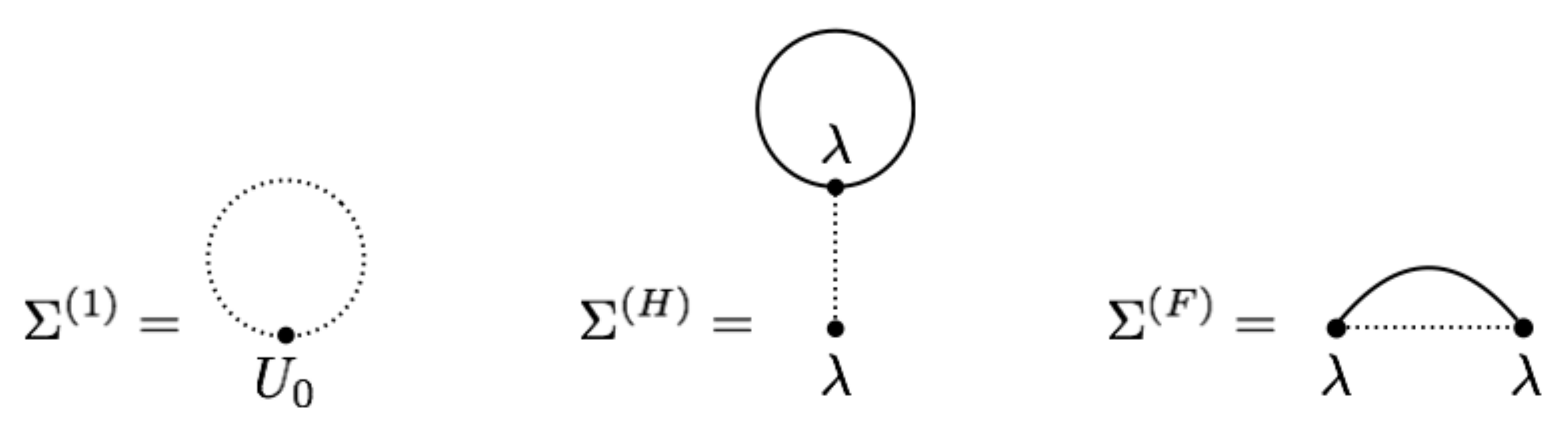}
\caption{Atomic self-energy insertions up to second order in the cavity field $a_{cl,q}$. The solid lines represent the bare (matrix) atom propagator while the dotted ones the bare (matrix) photon propagator. The coupling constants are $U_0=g_0^2/\Delta_{\rm a}$ and $\lambda=g_0\Omega/\Delta_{\rm a}$. In a double expansion in $N$ and sufficiently far from superradiant threshold, 
these diagrams are evaluated with bare photon propagators (see appendix~\ref{app:self}).}
\label{fig:self_energy_atoms}
\end{figure}

The possible Feynman diagrams resulting from this truncation are shown in Fig.~\ref{fig:self_energy_atoms}. 
As discussed in the key results section \ref{key_atoms}, it is important to note that this set of diagrams with the bare photon propagators would fail close to the superradiant threshold, where the corrections to the bare photon dynamics due to the atomic medium are important, ultimately leading to the softening of the collective modes. As described in Appendix~\ref{app:self}) , this forces us to include the polarization corrections to the photon propagator (see Sec.~\ref{subsec:photons}) which in turn functionally depend on the atomic distribution itself. We do not include these correction in the following quantum kinetic equation.

From the contribution to $\Phi$ linear in $V_{cl,q}$ we get
\[
\Sigma_{ab}^{(1)}(x,x^\prime)=\delta(x-x^\prime)\sum_\alpha \gamma_{ab}^\alpha\langle V_\alpha(x)\rangle\;,
\]
so that
\begin{align}
\label{sigma_1}
\Sigma^{(1)^{R}}(x,x^\prime)&=\delta(x-x^\prime)\frac12 U_0\eta_c^2(\mathbf{x})\left(\langle a_{cl}^*(t) a_{cl}(t)\rangle+\underbrace{\langle a_{q}^*(t) a_{q}(t)\rangle}_{=0\text{ causality}}\right)=\delta(x-x^\prime)\frac12 U_0\eta_c^2(\mathbf{x})\frac{i}{2}(g_{0,2x2}^K(t,t))_{11}\;.
\end{align}
This self-energy is real-valued. The Keldysh contraction of this term vanishes
\begin{align}
\Sigma^{(1)^{K}}(x,x^\prime)=\delta(x-x^\prime)\frac12 U_0\eta_c^2(\mathbf{x})\left(\underbrace{\langle a_{cl}^*(t) a_{q}(t)\rangle}_{=i(g_{0,2x2}^R(t,t))_{11}/2}+\underbrace{\langle a_{q}^*(t) a_{cl}(t)\rangle}_{=i(g_{0,2x2}^A(t,t))_{11}/2}\right)=0\;,
\end{align}

where we used the causality property $G^R(t,t)+G^A(t,t)=0$, valid for for any Green Funcion $G$.

From the terms quadratic in $V_{cl,q}$ we get the Hartree and Fock contractions shown 
in Fig.~\ref{fig:self_energy_atoms}
\begin{align}
\Sigma_{ab}^{(2)}(x,x^\prime)=&
-\delta(x-x^\prime)\int dy \sum_{\alpha,\beta}\sum_{a^\prime b^\prime} \gamma_{a b}^\beta\langle V_\beta(x) 
V_\alpha(y)\rangle\gamma_{a^\prime b^\prime}^\alpha G_{b^\prime a\prime}(y,y)
+\sum_{\alpha,\beta}\sum_{a^\prime b^\prime} \gamma_{a a^\prime}^\alpha G_{a\prime b^\prime }(x,x^\prime)\langle V_\alpha(x) V_\beta(x^\prime) \rangle\gamma_{b^\prime b}^\beta\nonumber\\
\equiv&\Sigma_{ab}^{(H)}(x,x^\prime)+\Sigma_{ab}^{(F)}(x,x^\prime)\;.
\end{align}
The Hartree contribution 
\begin{align}
\label{sigma_H}
\Sigma^{(H)^{R}}(x,x^\prime)&=-\delta(x-x^\prime)\frac12 \lambda^2\eta_{PC}(\mathbf{x})\int dy\;\eta_{PC}(\mathbf{y})
G^K(y,y)\left(\langle a_{cl}^*(x_t)+a_{cl}(x_t)\right) \left(\langle a_{q}^*(y_t)+a_{q}(y_t)\right)\rangle\nonumber\\
&=-\delta(x-x^\prime)\frac12 \lambda^2\eta_{PC}(\mathbf{x})\int dy\;\eta_{PC}(\mathbf{y})G^K(y,y)
\sum_{\ell,m=1}^2\frac{i}{2}(g_{0,2x2}^R(x_t,y_t))_{\ell m}\;,
\end{align}
is also real-valued, where $G^K$ depends on $F$ according to Eq.~(\ref{GK_par}) and we abbreviated for the 
mode functions $\eta_{PC}(\mathbf{x})\equiv\eta_p(\mathbf{x})\eta_c(\mathbf{x})$.
Again, the Keldysh component vanishes here $\Sigma^{(H)^{K}}(x,x^\prime)=0$.
%

We finally compute the Fock contributions to the self-energy
\begin{align}
&\Sigma^{(F)^{R}}(x,x^\prime)=G^R(x,x^\prime)\langle V_{cl}(x) V_{cl}(x^\prime)\rangle
+G^K(x,x^\prime)\langle V_{cl}(x) V_{q}(x^\prime)\rangle\;,\\
&\Sigma^{(F)^{K}}(x,x^\prime)=G^K(x,x^\prime)\langle V_{cl}(x) V_{cl}(x^\prime)\rangle
-\left(G^R(x,x^\prime)-G^A(x,x^\prime)\right)\left(\langle V_{q}(x) V_{cl}(x^\prime)\rangle
-\langle V_{cl}(x) V_{q}(x^\prime)\rangle\right)\;,
\nonumber
\end{align}
where the retarded contribution is now complex-valued and related by complex conjugation to the advanced component.
Inserting the explicit form of $V_{cl,q}$ yields

\begin{align}
\label{sigma_F}
\Sigma^{(F)^{R}}(x,x^\prime)&=\frac{i}{2}\lambda^2\eta_{PC}(\mathbf{x})\eta_{PC}(\mathbf{x}^\prime)\left[G^R(x,x^\prime) \frac12\sum_{\ell,m=1}^2(g_{0,2x2}^K(t,t^\prime))_{\ell m}+G^K(x,x^\prime) \frac12\sum_{\ell,m=1}^2(g_{0,2x2}^A(t,t^\prime))_{\ell m}\right]\\
\Sigma^{(F)^{K}}(x,x^\prime)&=\frac{i}{2}\lambda^2\eta_{PC}(\mathbf{x})\eta_{PC}(\mathbf{x}^\prime)\left[G^K(x,x^\prime) \frac12\sum_{\ell,m=1}^2(g_{0,2x2}^K(t,t^\prime))_{\ell m}+\right.\nonumber\\
&\left.+\mathrm{Im}[G^R(x,x^\prime)]\sum_{\ell,m=1}^2\left((g_{0,2x2}^R(t,t^\prime))_{\ell m}-(g_{0,2x2}^A(t,t^\prime))_{\ell m}\right)\right]
\nonumber\;.
\end{align}
with $\Sigma^{(F)^{A}}(x,x^\prime)=\Sigma^{(F)^{R}}(x,x^\prime)^*$.
%
%


\subsection{Quantum kinetic equation}
\label{sec:qke}

In order to derive the QKE equation it is convenient to work with the \emph{Wigner Transform} (WT) of the self-energies and propagators. We define therefore the WT of a two point function $A(x,x^\prime)$ as 
\[
\tilde{A}(X,p)=\int d\xi \;e^{-ip\cdot \xi}A(X+\frac{\xi}{2},X-\frac{\xi}{2})\;,\text{and }A(x,x^\prime)=\frac{1}{V}\sum_p e^{ip(x-x^\prime)}\tilde{A}(\frac{x+x^\prime}{2},p)\;,
\]
where $V$ is the volume of the system and we defined $p\cdot\xi\equiv \mathbf{p}\cdot\boldsymbol{\xi}-p_t\xi_t$, $\sum_p\equiv \sum_{\mathbf{p}}\int dp_t/2\pi$. In the following, we will exploit the useful properties
:
\begin{align}
\tilde{AB}=&\frac{1}{V}\sum_q\tilde{A}(X,p-q)\tilde{B}(X,q)\;,\nonumber\\
\tilde{A\circ B}=&\tilde{A}\tilde{B}+\frac{i}{2}\left(\partial_X\tilde{A}\partial_p\tilde{B}-\partial_p\tilde{A}\partial_X\tilde{B}\right)+\dots\;,
\end{align}
where the second equation corresponds to the ``slow'' approximation, where we assume that the dependence on the relative coordinate $\xi$ is much faster than the one on the absolute coordinate $X$ (for translationally invariant functions we only need to take the leading term). 
In terms of these slowly varying Wigner transformed self-energies, the Dyson Eq.~(\ref{dyson_RA}) 
reads
\begin{equation}
\label{dyson_WT}
\tilde{G}^{R(A)}(X,p)\simeq\frac{1}{p_t-\epsilon_{\mathbf{p}}-\tilde{\Sigma}^{R(A)}(X,p)}\;,
\end{equation}
and the parametrization (\ref{GK_par}) becomes
\begin{equation}
\label{GK_par_WT}
\tilde{G}^K(X,p)\simeq 2i\tilde{F}(X,p)\mathrm{Im}[\tilde{G}^R(X,p)]\;.
\end{equation}
The Dyson equation \eqref{BE} can now be rewritten as a \emph{Quantum Kinetic Equation} (QKE) for the steady state distribution $\tilde{F}(\mathbf{X},p)$ 
\begin{align}
\label{QKE} 
\bigg\{\mathbf{\nabla}_{\mathbf{p}}\big(\epsilon_{\mathbf{p}}+\mathrm{Re}\tilde{\Sigma}^R(\mathbf{X},p)\big)\cdot \mathbf{\nabla}_{\mathbf{X}}-\mathbf{\nabla}_{\mathbf{X}}\big(\mathrm{Re}\tilde{\Sigma}^R(\mathbf{X},p)\big)\cdot \mathbf{\nabla}_{\mathbf{p}}\bigg\}\tilde{F}(\mathbf{X},p)=I_{coll}[\tilde{F}]\;,
\end{align}
with the \emph{Collisional Integral}
\begin{align}
\label{coll_int} 
I_{coll}[\tilde{F}]=i\tilde{\Sigma}^K(\mathbf{X},p)+2\tilde{F}(\mathbf{X},p)\mathrm{Im}\tilde{\Sigma}^R(\mathbf{X},p)\;.
\end{align}
It is important to note that the only contribution to the collisional integral comes from the Fock self-energy (Eqs.~(\ref{sigma_F_WT_R}),(\ref{sigma_F_WT_K})). Various general expression of the  Wigner transformed self-energies appearing in Eq.~(\ref{QKE}) are 
collected in Appendix \ref{app:self}.


\subsection{Solving the quantum kinetic equation in thermodynamic limit $N\rightarrow \infty$}
\label{sec:atomic_steady_TL}

Because the photons can carry only an externally fixed momentum, 
only the Hartree contribution in Fig.~\ref{fig:self_energy_atoms} to the atomic self-energy survives in the TL, since i) $\Sigma^{(1)}$ is proportional to $U_0\propto 1/V$ but contains no atom propagator, ii) $\Sigma^{(F)}$ is proportional to $\lambda^2\propto 1/V$, contains one atom propagator but its momentum is fixed by momentum conservation, iii) $\Sigma^{(H)}$ is proportional to $\lambda^2\propto 1/V$ and contains one atom propagator whose momentum is not fixed by momentum conservation and therefore compensates the scaling of the coupling constant to give something finite in the TL. The full self-energy in the TL is thus
$\Sigma^R(x,x^\prime)=\Sigma^{(H)^{R}}(x,x^\prime)$,
$\Sigma^A(x,x^\prime)=\Sigma^{(H)^{R}}(x,x^\prime)$, 
and $\Sigma^K(x,x^\prime)=0$.
Therefore, in the TL \emph{no collisional integral is present} and only terms proportional to the derivatives of $\Sigma^{(H)}\in\mathbb{R}$ appear.


Using the explicit expression for the Hartree self-energy in the Appendix Eq.~(\ref{sigma_H_WT_final}) with 
the fact that quasiparticles are not broadened in the TL,
$\mathrm{Im}\tilde{G}^R(\mathbf{X},p)=-\pi\delta(p_t-\epsilon_{\mathbf{p}})$, 
we get the purely real
\begin{equation}
\label{sigma_H_WT_NSC}
\tilde{\Sigma}^{R}(\mathbf{X})=\frac{\lambda^2}{2}\cos(\mathbf{Q}\cdot\mathbf{X})\frac{\delta_c}{\delta_c^2+\kappa^2}\int\frac{d\mathbf{X}^\prime}{V}\sum_{\mathbf{q}}\tilde{F}(\mathbf{X}^\prime;\epsilon_{\mathbf{q}},\mathbf{q}) \cos(\mathbf{Q}\cdot\mathbf{X}^\prime)\;.
\end{equation}
%
the QKE becomes
\begin{align}
\label{QKE_TL_NSC} 
\frac{\mathbf{p}}{m}\cdot \mathbf{\nabla}_{\mathbf{X}}\tilde{F}(\mathbf{X},p)-\frac{\delta_c \lambda^2}{\delta_c^2+\kappa^2}\bigg(\int\frac{d\mathbf{X}^\prime}{V}\sum_{\mathbf{k}}\cos(\mathbf{Q\cdot\mathbf{X}^\prime})\tilde{F}(\mathbf{X}^\prime;\epsilon_{\mathbf{k}},\mathbf{k})\bigg)(\mathbf{\nabla}_{\mathbf{X}}\cos(\mathbf{Q\cdot\mathbf{X}}))\cdot \mathbf{\nabla}_{\mathbf{p}}\tilde{F}(\mathbf{X},p)=0\;,
\end{align}
which with $\tilde{F}(\mathbf{X};\epsilon_\mathbf{p},\mathbf{p})=1-2n_F(\mathbf{X},\mathbf{p})$ (note that this holds only if the 
quasiparticles are well defined), becomes
\begin{align}
\label{QKE_TL_NSC_population} 
\frac{\mathbf{p}}{m}\cdot \mathbf{\nabla}_{\mathbf{X}}n_F(\mathbf{X},\mathbf{p})-\frac{2\delta_c \lambda^2}{\delta_c^2+\kappa^2}\bigg(\int\frac{d\mathbf{X}^\prime}{V}\sum_{\mathbf{k}}\cos(\mathbf{Q\cdot\mathbf{X}^\prime})n_F(\mathbf{X}^\prime,\mathbf{k})\bigg)\sin(\mathbf{Q\cdot\mathbf{X}})\mathbf{Q}\cdot \mathbf{\nabla}_{\mathbf{p}}n_F(\mathbf{X},\mathbf{p})=0\;.
\end{align}
Eq.~(\ref{QKE_TL_NSC_population}) is a Vlasov equation identical to the one employed in \cite{griesser_vlasov,niedenzu_2011} to describe classical particles inside a transversally driven single-mode resonator. We have derived the latter from the most general miscroscopic quantum field theory of non equilibrium, where we know that the function $n_F(\mathbf{X},\mathbf{p})$ is the average number of particles with momentum $\mathbf{p}$ at position $\mathbf{X}$ in a non-translationally invariant system, treated in the slow approximation introduced above. Moreover, we have shown that the Vlasov equation (\ref{QKE_TL_NSC_population}) is valid also for quantum particles with fermionic statistics, provided we are in the TL.

As discussed in \cite{griesser_vlasov,niedenzu_2011}, an essential feature of the above equation is that any translationally invariant occupation $n_F(\mathbf{X},\mathbf{p})=n_F(\mathbf{p})$ is a solution, which means that if the atoms are initially in such a distribution, like the equilibrium Fermi distribution $n_F^{eq}(\epsilon_{\mathbf{p}})=(\exp\beta(\epsilon_{\mathbf{p}}-\mu)+1)^{-1}$, nothing will happen. Only upon including fluctuations this scenario is modified and a kind of thermalization takes place leading toward a steady state distribution which is different from the initial one \cite{niedenzu_2011}. This fluctuations are however negligible in the TL where the Vlasov Eq.~(\ref{QKE_TL_NSC_population}) becomes exact. However, it can be that these spatially inhomogeneous fluctuations are unstable and grow exponentially, giving rise to self-organization above a certain threshold $\lambda_{sr}$. 
We now want to derive this instability condition from the Vlasov equation (\ref{QKE_TL_NSC_population}), as 
done in \cite{griesser_vlasov} for classical particles.

\subsubsection{Self-organization threshold}

We begin by linearizing Eq.~(\ref{QKE_TL_NSC_population}) about the initial stationary equilibrium solution $n_F^{eq}(\epsilon_{\mathbf{p}})$:
\begin{equation}
n_F(\mathbf{X},\mathbf{p})=n_F^{eq}(\epsilon_{\mathbf{p}})+\delta n_F(\mathbf{X},\mathbf{p})\;,
\end{equation}
so that the equation becomes in one spatial dimension ($\mathbf{p}=p$, $\mathbf{X}=X$ )
\begin{align}
\label{QKE_vlasov_lin} 
\frac{p}{m}\partial_X \delta n_F(X,p)-\frac{2\delta_c \lambda^2}{\delta_c^2+\kappa^2}\bigg(\int\frac{dX^\prime}{V}\sum_{k}\cos(QX^\prime) \delta n_F(X^\prime,k)\bigg)\sin(QX)Q\partial_p n_F^{eq}(\epsilon_p)=0\;.
\end{align}
Now, with the aid of the FT $\delta n_F(X,p)=(1/\sqrt{V})\sum_P\exp(iPX) \delta n_F(P,p)$ and multiplying the equation by $\int (dX/\sqrt{V})\exp(-iPX)$ we get
\[
i\frac{p}{m}P \delta n_F(P,p)-\frac{\delta_c \lambda^2}{\delta_c^2+\kappa^2}Q(\partial_p n_F^{eq}(\epsilon_p))\sum_k\big(\delta n_F(Q,k)+\delta n_F(-Q,k)\big)\frac{1}{2i}\big(\delta_{P,Q}-\delta_{P,-Q}\big)=0\;.
\]
By adding the equation for $P=+Q$ with the one for $P=-Q$, summing over $p$ and defining $\sum_p\delta n_F(\pm Q,p)=\delta n_F(\pm Q)$ we get
\[
0=\bigg(\delta n_F(Q,k)+\delta n_F(-Q,k)\bigg)\bigg[1+\lambda^2\frac{\delta_c }{\delta_c^2+\kappa^2}\sum_p\frac{\partial_p n_F^{eq}(\epsilon_p)}{p/m}\bigg]\;.
\]
By requiring the latter to have a non trivial solution, i.e. the term between square bracket to vanish, we get the threshold coupling
\[
\lambda_{sr}^2=\frac{\delta_c^2+\kappa^2}{\delta_c\Pi_{slow}^R(\omega=0,Q)}\;,
\]
with the polarization propagator 
\begin{equation}
\Pi_{slow}^R(\omega=0,Q)=-\sum_p\frac{\partial_p n_F^{eq}(\epsilon_p)}{p/m}\;,
\end{equation}
which is nothing else but the polarization propagator (\ref{polariz_ret}) in the approximation $\epsilon_Q\ll \mu \overset{T=0}=\epsilon_F$, where
\begin{align}
n_F^{eq}(\epsilon_{p+Q})-n_F^{eq}(\epsilon_p)\simeq Q\partial_p n_F^{eq}(\epsilon_p)\;\text{ and }\;\; \epsilon_p-\epsilon_{p+Q}\simeq -\frac{pQ}{m}\;.
\label{eq:slow}
\end{align}
The regime $\epsilon_Q\ll \mu$ is indeed consistent with the slow approximation on which the derivation of the Vlasov Eq.~(\ref{QKE_TL_NSC_population}) is based, and which corresponds namely to the energy scale of the relative motion being much larger than the one of the center of mass motion set by $Q$. 
Therefore, we get a threshold which agrees with Eq.~(\ref{threshold_photon}) in the expected regime 
.

\subsection{Solving the quantum kinetic equation to order $1/N$}
\label{sec:one_over_N_QKE}

The next-to-leading order corrections to the QKE in the TL come from the Fock contributions to the self-energy, 
Eqs.~\eqref{sigma_F_WT_R},\eqref{sigma_F_WT_K}. In accordance with the approximations 
discussed above, we will use the bare photon propagator and also neglect the self-energy correction $\Sigma^{(1)}$, proportional to $U_0$.
Up to $1/N$, the self-energies appearing in the QKE are
%
%
$\tilde{\Sigma}^R(\mathbf{X},p)=\tilde{\Sigma}^{(H)^R}(\mathbf{X})+\tilde{\Sigma}^{(F)^R}(\mathbf{X},p)$
and
$\tilde{\Sigma}^K(\mathbf{X},p)=\tilde{\Sigma}^{(F)^K}(\mathbf{X},p)$.
%
such that the full QKE reads
\begin{align}
\label{QKE_one_overN} 
\left(\frac{\mathbf{p}}{m}+\nabla_{\mathbf{p}}\mathrm{Re}\tilde{\Sigma}^{(F)^R}(\mathbf{X},p)\right)\cdot \mathbf{\nabla}_{\mathbf{X}}\tilde{F}(\mathbf{X},p)-\mathbf{\nabla}_{\mathbf{X}}\left(\mathrm{Re}\tilde{\Sigma}^{(H)^R}(\mathbf{X})+\mathrm{Re}\tilde{\Sigma}^{(F)^R}(\mathbf{X},p)\right)\cdot \mathbf{\nabla}_{\mathbf{p}}\tilde{F}(\mathbf{X},p)=I_{coll}[\tilde{F}]\;,
\end{align}
with the collisional integral
\begin{align}
\label{coll_int_one_overN} 
I_{coll}[\tilde{F}]=i\tilde{\Sigma}^{(F)K}(\mathbf{X},p)+2\tilde{F}(\mathbf{X},p)\mathrm{Im}\tilde{\Sigma}^{(F)^R}(\mathbf{X},p)\;.
\end{align}
%
The Hartree SE is given in Eq.~(\ref{sigma_H_WT_NSC}) and is real. The Fock SE has instead also an imaginary part which gives rise to the collisional integral.
Restricting to the \emph{on-shell part}, we get
\begin{align}
&\tilde{\Sigma}^{(F)^R}(\mathbf{X},\epsilon_{\mathbf{p}},\mathbf{p})=\frac{\lambda^2}{4}\cos(2\mathbf{Q}\cdot\mathbf{X})\left[-\frac{\delta_c F_{\mathbf{p}}(\mathbf{X})}{\delta_c^2+\kappa^2}-\frac{i\kappa}{\delta_c^2+\kappa^2}\right]\nonumber\\
&+\frac{\lambda^2}{8}\sum_{\mathbf{q}=\pm\mathbf{Q}}\left[-\frac{\delta_c F_{\mathbf{p}+\mathbf{q}}(\mathbf{X})}{\delta_c^2+(\kappa-i(\epsilon_{\mathbf{p}}-\epsilon_{\mathbf{p}+\mathbf{q}}))^2}+\frac12\left(\frac{\epsilon_{\mathbf{p}}-\epsilon_{\mathbf{p}+\mathbf{q}}-\delta_c-i\kappa}{(\epsilon_{\mathbf{p}}-\epsilon_{\mathbf{p}+\mathbf{q}}-\delta_c)^2+\kappa^2}+\frac{\epsilon_{\mathbf{p}}-\epsilon_{\mathbf{p}+\mathbf{q}}+\delta_c-i\kappa}{(\epsilon_{\mathbf{p}}-\epsilon_{\mathbf{p}+\mathbf{q}}+\delta_c)^2+\kappa^2}\right)\right]\;,
\end{align}
with the on-shell distribution $F_{\mathbf{p}}(\mathbf{X})\equiv F(\mathbf{X};\epsilon_{\mathbf{p}},\mathbf{p})$.
Its real part reads
%
\begin{align}
\label{WT_sigma_F_Re}
\mathrm{Re}\tilde{\Sigma}^{(F)^R}(\mathbf{X},\epsilon_{\mathbf{p}},\mathbf{p})=&\frac{\lambda^2}{4}\cos(2\mathbf{Q}\cdot\mathbf{X})\frac{\delta_c F_{\mathbf{p}}(\mathbf{X})}{\delta_c^2+\kappa^2}
\nonumber\\
&+\frac{\lambda^2}{8}\sum_{\mathbf{q}=\pm\mathbf{Q}}\Bigg(\frac{-\delta_cF_{\mathbf{p+q}}(\mathbf{X})\big(\delta_c^2+\kappa^2-\omega_{\mathbf{q}}^2(\mathbf{p})\big)-\omega_{\mathbf{q}}(\mathbf{p}) \big(-\delta_c^2+\kappa^2+\omega_{\mathbf{q}}^2(\mathbf{p})\big)}{\big|\delta_c^2+(\kappa+i \omega_{\mathbf{q}}(\mathbf{p}))^2 \big|^2}\Bigg)\;,
\end{align}
with the particle-hole dispersion
\[
\omega_{\mathbf{q}}(\mathbf{p})=\frac{\mathbf{q}}{2m}\cdot(\mathbf{q}+2\mathbf{p})
\]
satisfying $\omega_{-\mathbf{q}}(\mathbf{p})=\omega_{\mathbf{q}}(-\mathbf{p})$.
Its imaginary part is
\begin{align}
2\mathrm{Im}\tilde{\Sigma}^{(F)^R}(\mathbf{X},\epsilon_{\mathbf{p}},\mathbf{p})=&-\frac{\lambda^2}{2}\frac{\kappa}{\delta_c^2+\kappa^2}\cos(2\mathbf{Q}\cdot\mathbf{X})
-
\frac{\lambda^2}{2}\frac{\kappa}{2}\sum_{\mathbf{q}=\pm\mathbf{Q}}\left[\frac{\delta_c^2+\kappa^2+\omega_{\mathbf{q}}(\mathbf{p})^2-2\delta_c\omega_{\mathbf{q}}(\mathbf{p})F_{\mathbf{p}+\mathbf{q}}(\mathbf{X})}{\big|\delta_c^2+(\kappa-i \omega_{\mathbf{q}}(\mathbf{p}))^2 \big|^2}
\right]\;.
\end{align}
The other contribution to the collisional integral comes from the imaginary Keldysh component of the Fock contraction
\begin{align}
i\tilde{\Sigma}^{(F)^K}(\mathbf{X},\epsilon_{\mathbf{p}},\mathbf{p})=&\frac{\lambda^2}{4}\frac{2\kappa(\delta_c^2+\kappa^2)F_{\mathbf{p}}}{(\delta_c^2+\kappa^2)^2}\cos(2\mathbf{Q}\cdot\mathbf{X})+
\frac{\lambda^2}{8}\sum_{\mathbf{q}=\pm\mathbf{Q}}\left[\frac{2\kappa(\delta_c^2+\kappa^2+\omega_{\mathbf{q}}(\mathbf{p})^2)F_{\mathbf{p}+\mathbf{q}}-4\delta_c\kappa\omega_{\mathbf{q}}(\mathbf{p})}{\big|\delta_c^2+(\kappa-i \omega_{\mathbf{q}}(\mathbf{p}))^2 \big|^2}
\right]\;.
\end{align}

Collecting the full collisional integral, the $\cos(\mathbf{Q}\cdot\mathbf{X})$-dependent part of $2\mathrm{Im}\Sigma^R$ cancels the one of $i\Sigma^K$ and we are left with
\begin{align}
\label{collisional_integral}
I_{coll}[F]=-\frac{\lambda^2}{4}\sum_{\mathbf{q}=\pm\mathbf{Q}}
\frac{\kappa}{\big|\delta_c^2+(\kappa-i \omega_{\mathbf{q}}(\mathbf{p}))^2 \big|^2}
\Bigg\{
\left[\delta_c^2+\kappa^2+\omega_{\mathbf{q}}(\mathbf{p})^2 \right]
\left[F_{\mathbf{p}}(\mathbf{X})-F_{\mathbf{p}+\mathbf{q}}(\mathbf{X})\right]
-2\delta_c\omega_{\mathbf{q}}(\mathbf{p})\left[F_{\mathbf{p}}(\mathbf{X}) F_{\mathbf{p}+\mathbf{q}}(\mathbf{X})-1\right]\Bigg\}\;,
\end{align}
%

In order to crack this equation, we make the following ansatz for the distribution function 
\begin{align}
\tilde{F}_{\mathbf{p}}(\mathbf{X})=F_{\mathbf{p}}^{(0)}+\cos(\mathbf{Q}\cdot\mathbf{X})F_{\mathbf{p}}^{(1)}+\cos(2\mathbf{Q}\cdot\mathbf{X})F_{\mathbf{p}}^{(2)}\;,
\label{eq:ansatz}
\end{align}
dropping higher multiples of $\mathbf{Q}$ with $F^{(0)}\propto O(1)$ 
and $F^{(1),(2)}\propto O(1/N)$. Upon plugging in the ansatz Eq.~(\ref{eq:ansatz}) 
into the QKE (\ref{QKE_one_overN}) and consistently keeping terms to $1/N$, we get
%
%
%
\begin{align}
\label{QKE_one_overN_ansatz} 
&-\frac{\mathbf{p}}{m}\left(\sin(\mathbf{Q}\cdot\mathbf{X})F_{\mathbf{p}}^{(1)}+\sin(2\mathbf{Q}\cdot\mathbf{X})F_{\mathbf{p}}^{(2)}\right)\\
+&\frac{\lambda^2}{2}\frac{\delta_c}{\delta_c^2+\kappa^2}\left[\sin(\mathbf{Q}\cdot\mathbf{X})\int\frac{d\mathbf{X}^\prime}{V}\sum_{\mathbf{q}}F_{\mathbf{q}}^{(1)}\cos^2(\mathbf{Q}\cdot\mathbf{X}^\prime)-\sin(2\mathbf{Q}\cdot\mathbf{X})F_{\mathbf{p}}^{(0)}\right]\mathbf{Q}\cdot\nabla_{\mathbf{p}}F_{\mathbf{p}}^{(0)}=I_{coll}[F^{(0)}]\;,\nonumber
\end{align}
where, importantly, the collisional integral, because it is already of order $1/N$, contains only the homogeneous component $F^{(0)}$.
As anticipated, the QKE now only contains terms of order $1/N$. Three mutually independent components appear, proportional to $1,\cos(\mathbf{Q}\cdot\mathbf{X})$, and $\cos(2\mathbf{Q}\cdot\mathbf{X})$. Each of these gives thus rise to a separate equation: three eqs. for the three unknowns $F^{(0),(1),(2)}$. The most important observation is that the equation resulting from the homogeneous component of the QKE only contains $F^{(0)}$ and it reads
\begin{equation}
\label{QKE_averaged}
I_{coll}[F^{(0)}]=0\;.
\end{equation}
Once $F^{(0)}$ is known, the two remaining equations allow then to compute separately $F^{(1)}$ and $F^{(2)}$.
The above equation for $F^{(0)}$ is thus the QKE for the $1/N$ dynamics of the \emph{spatially-averaged distribution function}. 
We now consider Eq.~\eqref{QKE_averaged} for the spatially averaged distribution $F^{(0)}$ and require each of its $\mathbf{q}=\pm\mathbf{Q}$ components to \emph{separately vanish}, to get 
\begin{align}
0=\left[\delta_c^2+\kappa^2+\omega_{\mathbf{Q}}(\mathbf{p})^2\right]\left[F_{\mathbf{p}}^{(0)}-F_{\mathbf{p}+\mathbf{Q}}^{(0)}\right]-2\delta_c\omega_{\mathbf{Q}}(\mathbf{p}) \left[F_{\mathbf{p}}^{(0)} F_{\mathbf{p}+\mathbf{Q}}^{(0)}-1\right]\;,
\end{align}
where we took the QKE resulting from the $+\mathbf{Q}$ component of $I_{coll}$ (the other component gives rise to the same equation).
We further \emph{approximately} write $F_{\mathbf{p}}^{(0)}\simeq 1\pm 2 n_{\mathbf{p}}$, where the $+(-)$ corresponds to bosons(fermions). Indeed, one has in general
\[
1\pm 2n_{\mathbf{p}}=\int\frac{d\omega}{2\pi}i\tilde{G}^{K}(\omega,\mathbf{p})=-2\int\frac{d\omega}{2\pi}\mathrm{Im}\tilde{G}^R(\omega,\mathbf{p})F^{(0)}(\omega,\mathbf{p})\;,
\]
i.e. the above approximation is exact where the atom propagator is the bare one: $\mathrm{Im}\tilde{G}^R(\omega,\mathbf{p})=-\pi\delta(\omega-\epsilon_{\mathbf{p}})$. Otherwise one has in principle to perform a weighted frequency-integral involving also the off-shell part of the distribution function.

\subsubsection{Non-equilibrium distribution function}

With $F_{\mathbf{p}}^{(0)}\simeq 1\pm 2 n_{\mathbf{p}}$ and in the slow-approximation (we restrict to $d=1$ along the cavity axis):
\[
n_{p+Q}\simeq n_{p}+Qn_{p}'\;\;\text{   and   }\;\; \omega_Q(p)\simeq\frac{pQ}{m}\;,
\]
we get the following QKE
\begin{align}
0=\mp\left(\delta_c^2+\kappa^2+(\frac{pQ}{m})^2\right)n_p'\mp 4\delta_c\frac{p}{m}\left(n_p+Qn_p'/2\right)-4\delta_c\frac{p}{m}n_p\left(n_p+Qn_p'\right)\;.
\end{align}

Now, further neglecting the corrections $Qn_p'\ll n_p$ and restricting to the \emph{dilute regime} $n_p\ll 1$, where the quantum-statistical effects resulting from the quadratic terms in $n_p$ can be neglected, yields the equation
\begin{align}
\label{tsallis_eq}
\left(\delta_c^2+\kappa^2+(\frac{pQ}{m})^2\right)n_p'=-4\delta_c\frac{p}{m}n_p\;,
\end{align}
which has been derived in \cite{niedenzu_2011} for classical particles and is solved by the Tsallis distribution
\begin{align}
\label{tsallis_dist}
n_p^{(ts)}\propto\left(1+4\frac{E_R\epsilon_p}{\delta_c^2+\kappa^2}\right)^{-\frac{\delta_c}{E_R}}\;,
\end{align}
which tends in the limit $\delta_c\gg E_R$ to a Boltzmann distribution with the \emph{effective temperature}
\begin{equation}
\label{Teff_atoms}
k_BT_{\rm eff}^{\rm (at)}=\frac{\delta_c^2+\kappa^2}{4\delta_c}\;.
\end{equation}

We can now extend this result to the \emph{quantum-degenerate regime} in which terms 
proportional higher powers of the density are included,
\begin{align}
\label{Tspiazzis_eq}
\left(\delta_c^2+\kappa^2+(\frac{pQ}{m})^2\right)n_p'=-4\delta_c\frac{p}{m}\left(n_p\pm n_p^2\right)\;,
\end{align}
Note that the sign of $n_p^2$ squared terms depends on the (quantum) statistics of the atoms.
This equation can be solved by the non-equilibrium distribution function
\begin{align}
\label{Tspiazzis_dist}
n_p^{(qnt)}=\frac{1}{C\left(1+4\frac{E_R\epsilon_p}{\delta_c^2+\kappa^2}\right)^{\frac{\delta_c}{E_R}}\mp 1}\;,
\end{align}
where $C$ is an arbitrary $p-independent$ constant, fixed by the normalization condition $\int dp\; n_p =N$. The effects of quantum statistics are important when this constant is at most of order one. For $C\gg 1$ we enter the classical regime described by the Tsallis distribution (\ref{tsallis_dist}).

As plotted and discussed in the key results section \ref{key_atoms}, in the limit $\delta_c\gg E_R$ this distribution tends to a Bose(Fermi) distribution 
with the effective temperature \eqref{Teff_atoms} and chemical potential $\mu$ fixed by the constant $C=\exp(-\mu/\kappa_BT_{\rm eff})$. 

We can finally estimate the thermalization rate leading to the distribution (\ref{Tspiazzis_dist}) in the slow approximation. In the collisional integral (\ref{collisional_integral}), the first term in the curly brackets yields the derivative part in Eq.~(\ref{Tspiazzis_eq}), while the second one yelds the part proportional to $n_p,n_p^2$ and is therefore the one quantifying the speed of thermalization. From the collisional integral we can thus read out the rate
\[
\Gamma_{\rm th}(p)=\frac{\lambda^2\kappa\delta_c\omega_{\mathbf{Q}}(\mathbf{p})}{\big|\delta_c^2+(\kappa-i \omega_{\mathbf{q}}(\mathbf{p}))^2 \big|^2}\;,
\]
which, taking the particle-hole excitations to have the characteristic energy $E_R\ll \delta_c,\kappa$, leads to the momentum-independent rate given in Eq.~(\ref{thermalization_rate}).



%

\section{Dynamics of photons}
\label{subsec:photons}

In this section, we employ the Keldysh framework to discuss the self-organization transition from 
the viewpoint of the photons. We describe the nontrivial photon dynamics resulting from the interplay of Markov cavity decay with the dispersion and absorption properties of the fermionic medium.
We here focus on the thermodynamic limit in which we know from the preceding section 
that the atom distribution retains the equilibrium form.

\begin{figure}[t]
\includegraphics[width=60mm]{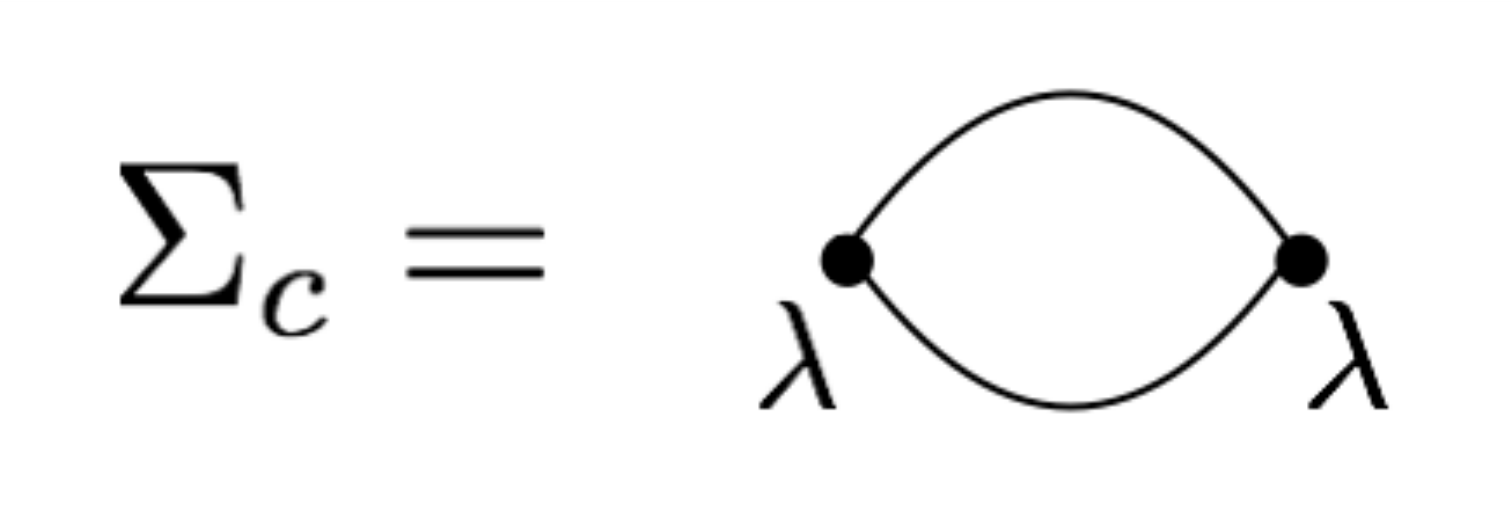}
\caption{
Bubble of atomic density fluctuations that damp the cavity photon.
}
\label{fig:bubble}
\end{figure}

We begin by integrating out the atomic degrees of freedom:
\begin{align}
Z=\frac{1}{\mathrm{Tr}[\hat{\rho}_0]}\int\mathcal{D}a^*\mathcal{D}a\;e^{i S_{0,\mathrm{C}}[a^*,a]}\det\left(i\underline{\mathbf{G}}_0^{-1}-i\underline{\mathbf{V}}\right)
=\underbrace{\frac{\det\left(i\underline{\mathbf{G}}_0^{-1}\right)}{\mathrm{Tr}[\hat{\rho}_0]}}_{=\mathrm{Tr}[\hat{\rho}_C]^{-1}}\int\mathcal{D}a^*\mathcal{D}a\;e^{i S_{0,\mathrm{C}}[a^*,a]}e^{\mathrm{Tr}\ln\left[1-\underline{\mathbf{G}}_0\cdot\underline{\mathbf{V}}\right]}\;,
\end{align}
which leads to the effective cavity-only action
\begin{align}
&S_{\rm eff}[a_{cl,q}^*,a_{cl,q}]=\int_{\infty}^\infty\frac{d\omega}{2\pi}\mathbf{a}^\dag(\omega)\cdot\left(\begin{array}{cc} 0 & \omega+\Delta_{\rm c}-i\kappa\\ \omega+\Delta_{\rm c}+i\kappa &2i\kappa\end{array}\right)\cdot\mathbf{a}(\omega)
-i\;\mathrm{Tr}\ln\left[1-\underline{\mathbf{G}}_0\cdot\underline{\mathbf{V}}\right]\;,
\label{cavity_only_action}
\end{align}
where we note that $\Delta_c < 0$ for a red-detuned pump. Expanding this to quadratic order in the fields $a_{cl}$, $a_q$, we get
%
%
\begin{equation}
\label{eq:expanded_log}
-i\;\mathrm{Tr}\ln\left[1-\underline{\mathbf{G}}_0\cdot\underline{\mathbf{V}}\right]=i\;\mathrm{Tr}\left[\underline{\mathbf{G}}_0\cdot\underline{\mathbf{V}}\right]+\frac{i}{2}\;\mathrm{Tr}\left[\underline{\mathbf{G}}_0\cdot\underline{\mathbf{V}}\cdot \underline{\mathbf{G}}_0\cdot\underline{\mathbf{V}}\right]
\;.
\end{equation}
We note that this "RPA" approximation is exact in the thermodynamic limit, while higher order terms exist as local vertices 
their impact on the photon dynamics is suppressed by powers of $1/N$ \cite{piazza_bose}. 
These terms are evaluated in App.~\ref{app:photons}. The photon 
propagator dressed by the atomic self-energies now reads
\begin{equation}
\label{photon_only_action_RPA}
S_{\rm RPA}[a_{cl,q}^*,a_{cl,q}]=\frac12\int_{-\infty}^\infty\frac{d\omega}{2\pi}\left(\begin{array}{cccc}a_{cl}^*(\omega) & a_{cl}(-\omega) & a_{q}^*(\omega) & a_{q}(-\omega)\end{array}\right)\cdot\left(\begin{array}{cc}0 & g_{2x2}^{A^{-1}}(\omega)\\ g_{2x2}^{R^{-1}}(\omega) & d_{2x2}^K(\omega)\end{array}\right) \cdot\left(\begin{array}{c}a_{cl}(\omega) \\ a_{cl}^*(-\omega) \\ a_{q}(\omega) \\ a_{q}^*(-\omega)\end{array}\right)\;,
\end{equation}
with $\delta_c = - \Delta_ c + 1/2 U_0 N$, 
\begin{equation}
\label{GR_cavity}
g_{2x2}^{R^{-1}}(\omega)\left(\begin{array}{cc}\omega-\delta_c+i\kappa+\Sigma_c^R(\omega) & \Sigma_c^R(\omega) \\ \Sigma_c^{R^{*}}(-\omega) & -\omega-\delta_c-i\kappa+\Sigma_c^{R^{*}}(-\omega)\end{array}\right)\;, 
\end{equation}
and $g_{2x2}^{A^{-1}}(\omega)=[g_{2x2}^{R^{-1}}(\omega)]^\dag$. 
The self-energies can be evaluated to be (App.~\ref{app:photons})
\begin{equation}
\label{self_energy_ret}
\Sigma_c^R(\omega)=\frac{g_0^2\Omega^2}{\Delta_{\rm a}^2}\frac12\Pi^R(\omega,\mathbf{Q})=\Sigma_c^{R^{*}}(-\omega)\;,
\end{equation}
where the retarded polarization bubble of Fig.~\ref{fig:bubble} 
is given by 
\begin{align}
\Pi^R(\omega,\mathbf{k})
=\sum_{\substack{\mathbf{k}^\prime}}\frac{n_{\rm F}(\epsilon_{\mathbf{k}+\mathbf{k}^\prime})-n_{\rm F}(\epsilon_{\mathbf{k}^\prime})}{\omega+\epsilon_{\mathbf{k}^\prime}-\epsilon_{\mathbf{k}+\mathbf{k}^\prime}+i0^+}\;,
\label{eq:Pi_ret}
\end{align}
$\Pi^A(\omega,\mathbf{k})=\Pi^R(\omega,\mathbf{k})^*$. The dressed, inverse Keldysh component of the photons reads 
\begin{equation}
d_{2x2}^K(\omega)\left(\begin{array}{cc}2i\kappa+\Sigma_c^K(\omega) & 0 \\ 0 & 2i\kappa+\Sigma_c^K(\omega) \end{array}\right)\;, 
\end{equation}
with
\begin{equation}
\label{self_energy_keldysh}
\Sigma_c^K(\omega)=\frac{g_0^2\Omega^2}{\Delta_{\rm a}^2}\frac12\Pi^K(\omega,\mathbf{Q})=-\Sigma_c^{K^{*}}(-\omega)\;.
\end{equation}
The Keldysh component of the bubble reads
%
\begin{align}
\Pi^K(\omega,\mathbf{k})&=\frac{2i \mathrm{Im}\Pi^R(\omega,\mathbf{k})}{1-2n_F(\omega)|_{\mu=0}}\;.
\end{align}
%

\subsection{Threshold behavior of the collective polariton modes}
\label{sec:photon_soft_mode}

The information about the quantum dynamics of the cavity mode is contained in the retarded Green function $g_{2x2}^{R}(\omega)$. In particular, the complex eigenenergies $E$ of the system are determined by its poles, satisfying the equation $\det g_{2x2}^{R^{-1}}(E)=0$:
\begin{equation}
\delta_{\rm c}^2-E^2+\kappa^2-2\delta_{\rm c}\mathrm{Re}\Sigma_c^R(E)-2i\delta_{\rm c}\left(\mathrm{Im}\Sigma_c^R(E)+\frac{E}{\delta_{\rm c}}\kappa\right)=0\;.
\label{poles}
\end{equation}
The equation above provides a useful insight regarding the issue of the competition between the dissipation resulting from the atomic bath $\mathrm{Im}\Sigma_c^R(E)$ and the Markov one resulting from the electromagnetic vacuum modes outside the cavity $\kappa$. 
The superradiant threshold is calculated by solving Eq.~(\ref{poles}) for $\lambda$ at $E=0$, and reads
\begin{equation}
\label{threshold_photon}
\lambda_{\rm sr}^2=\frac{\kappa^2+\delta_{\rm c}^2}{\delta_{\rm c}\mathrm{Re}\Pi^R(0,Q)}\;,
\end{equation}
where $\lambda=g_0\Omega/\Delta_{\rm a}$.

\begin{figure}[t]
\includegraphics[width=12cm]{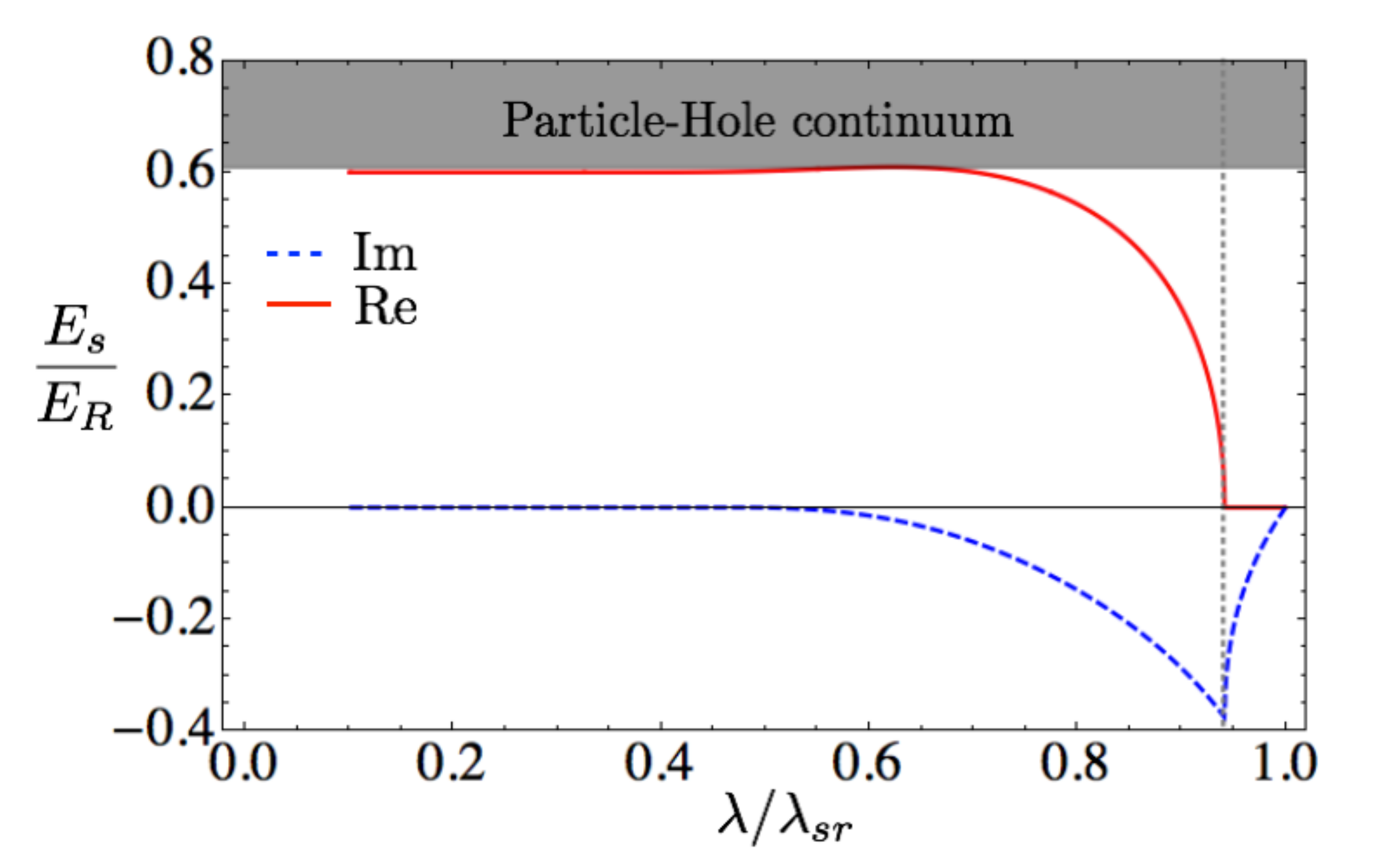}
\caption{Behavior of the soft polariton mode for a $d=1$ Fermi gas coupled to a cavity mode with $\delta_c=1.2\; E_R$ and $\kappa=1\; E_R$.}
\label{soft_behavior}
\end{figure}

We want now to calculate the soft mode complex energy $E_{\rm s}$ close and below threshold. 
Let us specify to the $d=1$ case, where for a non-perfect nesting we can assume that $\mathrm{Im}\Sigma_c^R(E_{\rm s})$ is \emph{identically zero} close enough to threshold. In this limt one can as well expand $\lambda^2\simeq\lambda_{\rm sr}^2+2\lambda_{\rm sr}(\lambda-\lambda_{\rm sr})$ and $\mathrm{Re}\Pi^R(E,Q)\simeq \mathrm{Re}\Pi^R(0,Q)+\mathcal{R}E^2$, with $\mathcal{R}$ some energy-independent function obtained by expanding the real part of Eq.~(\ref{eq:Pi_ret}). This gives a quadratic equation for $E$ with the lower solution being
\begin{equation}
\label{soft_mode_critical}
E_{\rm s}^{\rm 1d,fermi}=i\frac{\delta_{\rm c}\lambda_{\rm sr}^2 \mathrm{Re}\Pi^R(0,Q)}{\kappa}\frac{\lambda-\lambda_{\rm sr}}{\lambda_{\rm sr}}=i\frac{\delta_{\rm c}^2+\kappa^2}{\kappa}\frac{\lambda-\lambda_{\rm sr}}{\lambda_{\rm sr}}\;,
\end{equation}
which is purely imaginary negative(positive) below(above) threshold. This is exactly the same expression to be found for spins \cite{eth_soft,carmichael_2007,domokos_2008,dallatorre_2013}, and this is due to the fact that one-dimensional fermions provide a bath whose spectral density is non-zero only within the particle-hole continuum \cite{piazza_fermi}. 
For the Fermi gas in $d=1$, no solution to Eq.~(\ref{poles}) is found for frequencies whose real part lies within the particle-hole continuum, where the imaginary part of the photon self-energy is finite and frequency-independent. Therefore, for small $\lambda$ the real part of lower polariton eigenfrequency $E_s$ slowly comes out of the bottom of the particle-hole continuum and then vanishes where $E_s$ becomes purely imaginary, as discussed above. The eigenfrequency as a function of the coupling is depicted in Fig.~\ref{soft_behavior}.

It is interesting to compare the fermionic case above with a $d=3$ Bose gas \cite{piazza_bose}. Above $T_{BEC}$ we can safely neglect atom collisions and derive the photon self-energy in the same way described above. Close to threshold we have $2\mathrm{Im}\Sigma_c^R(E)\simeq(\lambda_{\rm sr}^2Q^3V/E_{\rm R}^2)f(T) E$ for small energies, with the dimensionless function $f(T)=(n\ell_T^3)^{-1}$ for $T\gg E_{\rm R}$ or $f(T)=e^{-E_{\rm R}/4T}$ for $T\ll E_{\rm R}$. The key difference with respect to the fermionic case is that the effective bath for the polariton mode provided by the Bose gas is frequency-dependent at low energies and vanishes linearly with frequency.
The polariton energy for a Bose gas close to threshold becomes
\begin{align}
\label{soft_mode_critical_bose}
E_{\rm s}^{\rm bose}=&i\frac{\delta_{\rm c}^2+\kappa^2}{\frac{\kappa^2+\delta_{\rm c}^2}{\mathrm{Re}\Pi(0,Q)}\frac{Q^3V}{E_{\rm R}^2}f(T)+\kappa}\frac{\lambda-\lambda_{\rm sr}}{\lambda_{\rm sr}}
\xrightarrow{\delta_{\rm c},\kappa\gg E_{\rm R},T}i\frac{\mathrm{Re}\Pi(0,Q) E_{\rm R}^2}{Q^3Vf(T)}\frac{\lambda-\lambda_{\rm sr}}{\lambda_{\rm sr}}\;,
\end{align}
where we see that $\kappa$ disappears from the prefactor in the bad-cavity and/or large detuning limit. 

\subsection{Effective temperature of photons}
\label{sec:photon_LET}

Given the  $(R,A,K)$ components of the effective cavity propagator, we can compute the Keldysh distribution function $F(\omega)$, which parametrizes the Keldysh Green function 
\begin{align}
\label{GK_cavity}
g_{2x2}^K(\omega)&=g_{2x2}^R(\omega)F_{2x2}(\omega)-F_{2x2}(\omega)g_{2x2}^A(\omega)
\leftrightarrow\;d_{2x2}^K(\omega)=g_{2x2}^{R^{-1}}(\omega)F_{2x2}(\omega)-F_{2x2}(\omega)g_{2x2}^{A^{-1}} (\omega)\;.
\end{align}
The above equation can be solved for the Hermitian matrix $F_{2x2}(\omega)$, for instance by mapping the latter together with $d_{2x2}^K(\omega)$ into $4x4$ vectors and solving the corresponding linear system.

For fermions in $d=1$, and away from perfect nesting $Q\neq 2k_{\rm F}$, since the imaginary part of $\Sigma_c(\omega)$ is \emph{identically zero} for small enough frequencies, we get
\[
F_{2x2}(\omega)\xrightarrow{\omega\to 0}\sigma_{z}+\frac{1}{\omega}\frac{\lambda^2}{2}\Pi^R(\omega,Q)\sigma_x\;,
\]
which, as found for spins \cite{dallatorre_2013}, is a traceless matrix with two opposite eigenvalues $f_{\pm}(\omega)$ which diverge like $2T_{\rm eff}/\omega$ with the low energy effective temperature (LET)
\begin{equation}
\label{effT_photon}
k_BT_{\rm eff}^{\rm (ph)}=\frac{1}{4}\lambda^2\mathrm{Re}\Pi^R(0,Q)\;,\text{  1d, fermi}\;,
\end{equation}
which does not involve $\kappa$. On the other hand, at perfect nesting $Q=2k_{\rm F}$ we have $\mathrm{Im}\Sigma_c^R(\omega)=\text{const.}$ for small frequencies. This has two consequences: first, the non identically vanishing imaginary part gives the matrix $F_{2x2}(\omega)$ a non-vanishing trace and thus two different eigenvalues $f_1\neq -f_2$; second, the frequency-independent imaginary part removes the $1/\omega$ divergence of the eigenvalues, leading to a \emph{non-thermal behavior}. In particular, in the bad-cavity limit of very large $\kappa$ we get
\[
f_1^{Q=2k_{\rm F}}\xrightarrow{\kappa\gg E_{\rm R},\omega\to 0}\frac{\sqrt{\mathrm{Re}\Sigma_c^{R^2}(0)+\mathrm{Im}\Sigma_c^{R^2}(0)}}{\mathrm{Im}\Sigma_c^{R}(0)}\xrightarrow{\mathrm{Im}\Sigma_c^{R}(0)\gg \mathrm{Re}\Sigma_c^{R}(0)}1\;.
\]
%
%

As comparison, let us instead consider a Bose-gas in $d=3$, whose behavior at low frequency we already discussed above in relation to the soft mode. In particular, since the imaginary part of the self-energy vanishes linearly with $\omega$, we get a non-vanishing trace of  $F_{2x2}(\omega)$ with eigenvalues
\begin{align}
f_{1,2}^{\rm bose}
&\xrightarrow{\omega\to 0,\kappa\gg E_{\rm R}}\pm\frac{1}{\omega}\frac{\lambda^2\mathrm{Re}\Pi^R(0,Q)}{\sqrt{4+\frac{\lambda^4Q^6V^2f^2(T)}{E_{\rm R}^4}}}\;,
\end{align}
so that the effective temperature becomes
\begin{align}
k_BT_{\rm eff}^{\rm (ph)}&=\frac{\lambda^2\mathrm{Re}\Pi^R(0,Q)}{\sqrt{4+\frac{\lambda^4Q^6V^2f^2(T)}{E_{\rm R}^4}}}\nonumber\\
&\to\left\{\begin{array}{cc} \frac{\lambda^2\mathrm{Re}\Pi^R(0,Q)}{4}& \;\;T\ll E_{\rm R} \\ \frac{E_{\rm R}^2/T}{2Q^3V}n\ell_T^3\propto T^{-5/2} & \;\;T\gg E_{\rm R}\end{array}\right.
\end{align}

\subsection{Cavity spectrum and linewidth narrowing for $\kappa\rightarrow \infty$}
\label{sec:spectral}

\begin{figure}
\includegraphics[width=165mm]{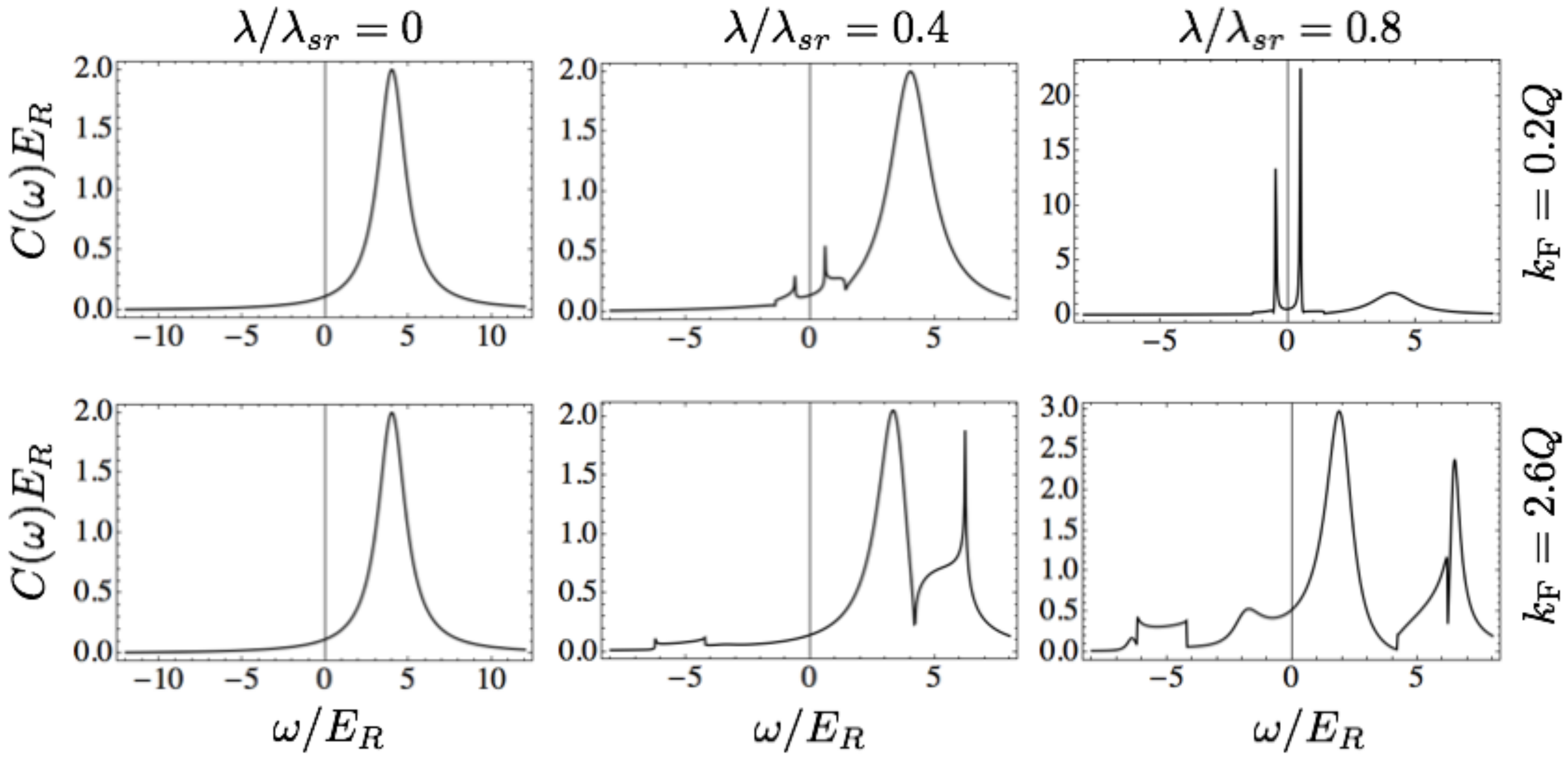}
\caption{
Comparison of the cavity photon correlation spectrum upon approaching 
threshold (from left to right) for small densities (top row) to large densities (bottom row).
The fermionic particle-hole continuum is visible between the polariton peaks in the bottom row.
(Sub-natural) Cavity-linewidth narrowing is most pronounced in the top row upon approaching threshold.
Parameters are $\kappa = 1 E_R$.
}
\label{correlation_typical}
\end{figure}

The cavity spectrum can be obtained from the Keldysh Green function
\begin{equation}
\label{correlation_function}
C(\omega)=i[g_{2x2}^K(\omega)]_{11}\;,
\end{equation}
which is the Fourier transform of the symmetrized correlation function $i[g_{2x2}^K(t)]_{11}=\langle\hat{a}(t)\hat{a}^\dag(0)+\hat{a}^\dag(0) \hat{a}(t)\rangle$, so that $\int d\omega C(\omega)=1+2\langle\hat{a}^\dag\hat{a}\rangle$. 
In Fig.~\ref{correlation_typical}, we discuss the spectrum for various $\lambda$.


One important aspect of the spectral response and correlation function introduced above is the linewidth of the light emitted. In this respect, it must be stressed that the cavity decay and the atomic one do not simply sum up to give rise to the width of the peaks to be seen in 
Fig.~\ref{correlation_typical}. In particular, in the experimentally relevant regime, where $\kappa$ is much larger than 
$E_{\rm R}$, the cavity and atomic dynamics decouple, and the system enters the bad-cavity limit. This regime is very interesting from the point of view of engineering narrow linewidth sources, as illustrated in Fig.~\ref{damping}, where the damping rate of the soft polariton mode is shown together with the spectral density correlation function $C(\omega)$, for the case of a $1d$ fermi gas. 
The damping rate is given by $\Gamma=-\mathrm{Im}[E_s]$ where $E_s$ is the (complex) eigenfrequency, solution of Eq.~(\ref{poles}), of the collective polariton mode which becomes soft at the superradiant threshold. As discussed in Sec.~\ref{sec:photon_soft_mode},close to threshold $E_s$ becomes purely imaginary exactly where the damping is maximum. After this point, the damping rate decreases and vanishes linearly with $\lambda_{sr}-\lambda$, as can be seen in Fig.~\ref{damping} consistently with Eq.~(\ref{soft_mode_critical}). From the inset of Fig.~\ref{damping} the role of the cavity decay $\kappa$ is apparent: the highest damping rate has a maximum for $\kappa\simeq E_R$ then decreases monotonously with $\kappa$. This behavior, which is restricted to values of $\lambda$ up to the point of maximum damping, can be analytically seen by considering the polariton eigenfrequency by solving Eq.~(\ref{poles}) in the limit of large $\kappa$. Outside the fermionic particle-hole continuum, the imaginary part of the soft polariton mode reads:
\[
\mathrm{Im}[E_s]\overset{\kappa\gg \delta_c,E_R}{\simeq}-\frac{E_R^2}{\kappa}\;.
\]
This behavior reflects onto the spectral density correlation function $C(\omega)$, as can be seen in Fig.~\ref{damping}, where $C(\omega)$ is shown both inside and outside the bad-cavity regime. By increasing $\kappa$ by a factor $10$, the lower polariton peak (and its partner at negative frequency as well) narrows visibly while the upper polariton peak around the cavity detuning $\delta_c$ is washed out.
In the limit of very large cavity decay, the spectral pureness of the soft polariton peak is ultimately spoiled only by absorption within the atomic medium. Fortunately, for the non-perfectly nested (and collisionless) Fermi gas, when the peak is outside the particle-hole continuum, atomic decay is absent (or exponentially supressed at finte $T$), leading to the very narrow linewidth apparent in Fig.~\ref{damping}.
It is clear the a larger $\kappa$ decreases the number of photons in the cavity, which is compensated by correspondingly increasing the pump strength and thereby $\lambda$. 

As clearly shown in Fig.~\ref{correlation_typical}, the narrowing of the soft mode in the bad-cavity limit takes place only in the regime where the cavity detuning $\delta_c$ is larger than the bottom of the particle-hole continuum, so that the soft mode is ``matter-like'' and emerges from the latter as described in Fig.~\ref{soft_behavior}. In the opposite regime the soft mode starts from the cavity frequency and, being ``photon-like'', is strongly affected by cavity decay.

%
%

\subsection{Squeezing in quadrature fluctuations}

After discussing the effect of dispersion and absorption on the spectral features we now consider what happens to the fluctuations in the quadratures of the intracavity light, which provides a measure of squeezing. The quadrature spectrum at a given angle $\theta$
\begin{equation}
\langle\hat{X}_\theta(\omega)\hat{X}_\theta(-\omega)\rangle=\frac{i}{\delta_{\rm c}}\left(\begin{array}{cc}e^{-i\theta}&e^{i\theta}\end{array}\right)g_{2x2}^K(\omega)\left(\begin{array}{c}e^{i\theta}\\e^{-i\theta}\end{array}\right)
\end{equation}
can be obtained from the Keldysh Green function. Upon integration over frequency we then compute the equal-time quadrature fluctuations
\begin{equation}
\langle\hat{X}_\theta(t)\hat{X}_\theta(t)\rangle=\int\frac{d\omega}{2\pi}\langle\hat{X}_\theta(\omega)\hat{X}_\theta(-\omega)\rangle\;,
\end{equation}
as depicted in Fig.~\ref{quadrature_large_density}.
Interestingly, while the optimal \emph{intra-cavity} squeezing for bosons and spins is always above the vacuum shot-noise \cite{dallatorre_2013}, the one for fermions in $d=1$ can go below the shot-noise value. The value of the corresponding squeezing is enhanced by increasing the density of the fermions thereby increasing the ratio $k_{\rm F}/Q$, as illustrated in the left panel of Fig.~\ref{quadrature_large_density}. This lowers the fluctuations down to at most $50\%$ squeezing, which is obtained at threshold $\lambda=\lambda_{\rm sr}$ and for $k_{\rm F}/Q\gg 1$, where the optimal angle then coincides with the one of spins/bosons. The reason for the squeezing can be traced back to a strongly flattened real part of the self energy at low frequencies due to the $1d$ nature of the fermi gas, as shown in the right panel of Fig.~\ref{quadrature_large_density}. This mimics the situation in a standard optical parametric oscillator, where a frequency-independent drive is needed in order to obtain squeezing. In the optical parametric oscillator, the squeezing can indeed reach intra-cavity values of $50\%$ of the vacuum shot-noise level \cite{carmichael_book}. 

\section{Conclusions}
%
In this work, we considered what is perhaps the driven-dissipative quantum system under best experimental control at the moment: an ultracold atomic gas coupled to a single-mode optical cavity. Despite its simplicity of being essentially a damped harmonic oscillator 
coupled to a non-interacting gas of particles, our results support the hypothesis that 
this system can become a fundamental testing ground for quantum statistical mechanics far-from-equilibrium. 
We found that Markov noise in combination with quantum statistics
generalizes the (equilibrium) 
Fermi-Dirac (and Bose-Einstein) to modified power-law distributions out-of-equilibrium with 
exponents given by the external optical parameters. 
The cavity spectrum shows narrow matter-light polaritonic sidebands in the bad cavity limit, 
whose width is only limited by atomic absorption, the latter being suppressed in the collisionless 
Fermi gas. 

Our results were derived in a continuum Keldysh description suitable for a large number 
of atoms, which may be helpful for future investigations of transport and slow dynamics close to the 
superradiance threshold.

After finalization of this work, we became aware of an unpublished computation 
of the atomic distribution function including quantum effects from a Master equation approach 
\cite{griesser_private}, consistent with our results.

\acknowledgments

We thank T. Griesser, G. Morigi, W. Niedenzu, H. Ritsch, A. Rosch, S. Sch\"utz, and W. Zwerger for 
discussions. This work was supported by the DFG under grant Str 1176/1-1, by the Leibniz 
prize of A. Rosch, by the NSF under Grant DMR-1103860, by the Templeton foundation, 
by the Center for Ultracold Atoms (CUA), and by the Multidisciplinary University 
Research Initiative (MURI). 

\appendix

\section{Atom self-energies}
\label{app:self}
We here collect the Wigner-transformed (WT) atom self-energies appearing in the quantum kinetic equation (\ref{QKE}).
The WT of $\Sigma^{(1)^R}$ is 
\[
\tilde{\Sigma}^{(1)^R}(X,p)=i\frac{U_0}{4}\eta_c^2(\mathbf{X})\int\frac{d\omega}{2\pi}(g_{0,2x2}^K(\omega))_{11}\;,
\]
where we assumed to be in the \emph{steady state}, so that $g_{0,2x2}(t,t^\prime)=g_{0,2x2}(t-t^\prime)$ is translationally invariant. Since the WT of a translationally invariant function is its Fourier transform, we used $g_{0,2x2}(\omega)$ without $\sim$.

After a similar calculation, we get
\begin{align}
\label{sigma_H_WT}
\tilde{\Sigma}^{(H)^R}(X,p)=+\lambda^2\eta_{PC}(\mathbf{X})\frac12\sum_{\ell,m=1}^2(g_{0,2x2}^R(\omega=0))_{\ell m}
\times\int\frac{d\mathbf{X}^\prime}{V}\sum_q\tilde{F}(\mathbf{X}^\prime,q)\mathrm{Im}[\tilde{G}^R(\mathbf{X}^\prime,q)]\eta_{PC}(\mathbf{X}^\prime)\;.
\end{align}



Note that $\mathrm{Im}[\tilde{G}^R(\mathbf{X}^\prime,q)]$ depends on $F$ and on $\Sigma^{R,A,K}$ itself through equation (\ref{dyson_WT}). The latter dependence implies in general a self-consistent calculation of $\Sigma^{R,A,K}$. For generality, 
we here 
also carry through the self-energy in the photon propagator (despite dropping it in the actual calculation).
From Eq.~(\ref{GR_cavity}) we get
\begin{equation}
\label{GcR}
\sum_{\ell,m=1}^2(g_{0,2x2}^R(\omega))_{\ell m}=\frac{-2\delta_c}{\delta_c^2+(\kappa-i\omega)^2-2\delta_c\Sigma_c^R(\omega)}\;,
\end{equation}
with the photon self-energy $\Sigma_c^R(\omega)$ also depending on $F$ and $\Sigma^{R,A,K}$. 
Starting back from Eq.~\eqref{eq:expanded_log} (see also appendix \ref{app:photons}), we can express the photon self-energy in terms of the WTs of $F$ and $G$ as
\begin{align}
\label{sigma_cavity_ret}
&\Sigma_c^R(\omega)=-\lambda^2\int\frac{d\mathbf{X}}{V}\sum_p\frac12\times\nonumber\\
\times&\left\{\cos(2\mathbf{Q}\cdot\mathbf{X})\left[\tilde{G}^R(\mathbf{X},p)\tilde{F}(\mathbf{X};p_t+\omega,\mathbf{p})\mathrm{Im}\tilde{G}^R (\mathbf{X};p_t+\omega,\mathbf{p})+\tilde{F}(\mathbf{X},p) \mathrm{Im}\tilde{G}^R (\mathbf{X},p)\tilde{G}^A(\mathbf{X};p_t+\omega,\mathbf{p}) \right]+\right.\nonumber\\
+&\left.\frac12 \sum_{\mathbf{q}=\pm\mathbf{Q}}\left[\tilde{G}^R(\mathbf{X},p)\tilde{F}(\mathbf{X};p_t+\omega,\mathbf{p}+\mathbf{q})\mathrm{Im}\tilde{G}^R (\mathbf{X};p_t+\omega,\mathbf{p}+\mathbf{q})+\tilde{F}(\mathbf{X},p) \mathrm{Im}\tilde{G}^R (\mathbf{X},p)\tilde{G}^A(\mathbf{X};p_t+\omega,\mathbf{p}+\mathbf{q}) \right]
\right\}\;,
\end{align}
where we neglected the pump momentum transfer so that
$\eta_{PC}(\mathbf{x})\simeq\cos(\mathbf{Q}\cdot\mathbf{x})$.
%

%

Upon substituting Eq.~(\ref{GcR}) into Eq.~(\ref{sigma_H_WT}) we can thus write
\begin{equation}
\label{sigma_H_WT_final}
\tilde{\Sigma}^{(H)^R}(\mathbf{X},p)=-\lambda^2\cos(\mathbf{Q}\cdot\mathbf{X})\frac{\delta_c}{\delta_c^2+\kappa^2 -2\delta_c\Sigma_c^R(0)
}\int\frac{d\mathbf{X}^\prime}{V}\sum_q\tilde{F}(\mathbf{X}^\prime,q)\mathrm{Im}[\tilde{G}^R(\mathbf{X}^\prime,q)]\cos(\mathbf{Q}\cdot\mathbf{X}^\prime)\;,
\end{equation}
where $\tilde{G}^R(\mathbf{X}^\prime,q)$ is given in Eq.~(\ref{dyson_WT}) and $\Sigma_c^R(\omega)$ in Eq.~(\ref{sigma_cavity_ret}).

A longer but analogous procedure yields the WT of the Fock contributions to the self-energy. First the retarded:
\begin{align}
\label{sigma_F_WT_R}
\tilde{\Sigma}^{(F)^R}(\mathbf{X},p)=&\lambda^2\frac{\cos(2\mathbf{Q}\cdot\mathbf{X})}{4}\int\frac{dq_t}{2\pi}\bigg\{\frac{2\delta_c\tilde{F}(\mathbf{X};q_t,\mathbf{p})\mathrm{Im}\tilde{G}^R(\mathbf{X};q_t,\mathbf{p})}{\delta_c^2+(\kappa-i(p_t-q_t))^2-2\delta_c\Sigma_c^R(p_t-q_t)}\nonumber\\
&+\frac{\left(\delta_c^2+\kappa^2+(p_t-q_t)^2\right)\left(2\kappa-i\Sigma_c^K(p_t-q_t)\right)\tilde{G}^R (\mathbf{X};q_t,\mathbf{p})}{\big|\delta_c^2+\left(\kappa-i(p_t-q_t)\right)^2-2\delta_c\Sigma_c^R(p_t-q_t)\big|^2} \bigg\}\nonumber\\
&
+\lambda^2\frac{1}{8}\sum_{\mathbf{q}=\pm\mathbf{Q}}\int\frac{dq_t}{2\pi}\bigg\{\frac{2\delta_c\tilde{F}(\mathbf{X};q_t,\mathbf{p}+\mathbf{q})\mathrm{Im}\tilde{G}^R(\mathbf{X};q_t,\mathbf{p}+\mathbf{q})}{\delta_c^2+(\kappa-i(p_t-q_t))^2-2\delta_c\Sigma_c^R(p_t-q_t)}+\nonumber\\
&+\frac{\left(\delta_c^2+\kappa^2+(p_t-q_t)^2\right)\left(2\kappa-i\Sigma_c^K(p_t-q_t)\right)\tilde{G}^R (\mathbf{X};q_t,\mathbf{p}+\mathbf{q})}{\big|\delta_c^2+\left(\kappa-i(p_t-q_t)\right)^2-2\delta_c\Sigma_c^R(p_t-q_t)\big|^2} \bigg\}\;,
\end{align}
where $\tilde{G}^R(\mathbf{X}^\prime,q)$ is given in Eq.~(\ref{dyson_WT}), $\Sigma_c^R(\omega)$ in Eq.~(\ref{sigma_cavity_ret}), and
\begin{align}
\label{sigma_cavity_kel}
\Sigma_c^K(\omega)=&i\frac{\lambda^2}{2}\int\frac{d\mathbf{X}}{V}\sum_p\frac12\times\nonumber\\
\times&\left\{\cos(2\mathbf{Q}\cdot\mathbf{X})4 \mathrm{Im}\tilde{G}^R(\mathbf{X},p) \mathrm{Im}\tilde{G}^R (\mathbf{X};p_t+\omega,\mathbf{p})\left[1-\tilde{F}(\mathbf{X},p) \tilde{F}(\mathbf{X};p_t+\omega,\mathbf{p})\right]+\right.\nonumber\\
+&\left. 2\mathrm{Im}\tilde{G}^R(\mathbf{X},p) \sum_{\mathbf{q}=\pm\mathbf{Q}}\mathrm{Im}\tilde{G}^R (\mathbf{X};p_t+\omega,\mathbf{p}+\mathbf{q})\left[1-\tilde{F}(\mathbf{X},p) \tilde{F}(\mathbf{X};p_t+\omega,\mathbf{p}+\mathbf{q})\right]
\right\}\;.
\end{align}
Finally the Keldysh component
\begin{align}
\label{sigma_F_WT_K}
\tilde{\Sigma}&^{(F)^K}(\mathbf{X},p)=i\lambda^2\frac{\cos(2\mathbf{Q}\cdot\mathbf{X})}{2}\int\frac{dq_t}{2\pi}
\times\mathrm{Im}\tilde{G}^R(\mathbf{X};q_t,\mathbf{p})
\\
&
\bigg\{\frac{\tilde{F}(\mathbf{X};q_t,\mathbf{p})\left(\delta_c^2+\kappa^2+(p_t-q_t)^2\right)\left(2\kappa-i\Sigma_c^K(p_t-q_t)\right)+4\delta_c\left(\kappa(p_t-q_t)+\delta_c\Sigma_c^R(p_t-q_t)\right)}{\big|\delta_c^2+\left(\kappa-i(p_t-q_t)\right)^2-2\delta_c\Sigma_c^R(p_t-q_t)\big|^2} \bigg\}\nonumber\\
&+i\frac{\lambda^2}{4}\int\frac{dq_t}{2\pi}\sum_{\mathbf{q}=\pm\mathbf{Q}}\times\mathrm{Im}\tilde{G}^R(\mathbf{X};q_t,\mathbf{p}+\mathbf{q})
\nonumber\\
&
\bigg\{\frac{\tilde{F}(\mathbf{X};q_t,\mathbf{p}+\mathbf{q})\left(\delta_c^2+\kappa^2+(p_t-q_t)^2\right)\left(2\kappa-i\Sigma_c^K(p_t-q_t)\right)+4\delta_c\left(\kappa(p_t-q_t)+\delta_c\Sigma_c^R(p_t-q_t)\right)}{\big|\delta_c^2+\left(\kappa-i(p_t-q_t)\right)^2-2\delta_c\Sigma_c^R(p_t-q_t)\big|^2} \bigg\}\;,\nonumber
\end{align}
where again $\tilde{G}^R(\mathbf{X}^\prime,q)$ is given in Eq.~(\ref{dyson_WT}), $\Sigma_c^R(\omega)$ in Eq.~(\ref{sigma_cavity_ret}), and $\Sigma_c^K(\omega)$ in Eq.~(\ref{sigma_cavity_kel}).

The difference between bosons and fermions is only in the sign of the Hartree self-energy (\ref{sigma_H}) (due to the fermionic loop) and in the fact that $F=1+2n_B$ instead of $F=1-2n_F$.
Regarding the Hartree contribution, this implies that the fermionic loop sign difference between bosons and fermions is compensated by the sign difference in $F$ as a function of $n_{B/F}$. This means in turn that the Vlasov Eq.~(\ref{QKE_TL_NSC_population}) is exactly the same both for bosons and fermions.
On the other hand, since the Fock diagram (Fig.~\ref{fig:self_energy_atoms}) does not contain any loop, the corresponding self-energy Eqs.~(\ref{sigma_F_WT_R}),(\ref{sigma_F_WT_K}) is the same for bosons and fermions except the sign difference in $F=1\pm 2n_{B/F}$. This implies that the collisional integral (\ref{collisional_integral}) looks the same for both bosons and fermions, but, once written as a function of $n_{B/F}$, it is actually statistics-dependent.

In the $1/N$ expansion in the TL we adopted to obtain the approximated QKE given in \eqref{QKE_one_overN}, all the atomic propagators appearing in the above self-energies must be the bare ones \cite{polyakov_book}. The photon propagator instead should contain the self-energy corrections $\Sigma_c^{R,A,K}$. This corrections are very important close to threshold where the photon is strongly hybridized with the matter and becomes soft. In the derivation of our atomic QKE \eqref{QKE_one_overN} we however neglect this corrections assuming that the coupling $\lambda$ is sufficiently smaller than $\lambda_{sr}$.

\section{Photon self-energies}
\label{app:photons}

We here collect the self-energy terms of photons from Subsec.~\ref{subsec:photons}. 
Upon transforming Eq.~(\ref{eq:expanded_log}) to the space-time domain, the first term reads
\begin{align}
\mathrm{Tr}\left[\underline{\mathbf{G}}_0\cdot\underline{\mathbf{V}}\right]
V_{q}(t)\delta(t-t^\prime)
=\int_\infty^\infty\!\!dt\left[\underbrace{\left(G_0^A(t,t)+G_0^R(t,t)\right)}_{=0}V_{cl}(t)+G_0^K(t,t)V_{q}(t)\right]
=\int_\infty^\infty\!\!dt\;G_0^K(t,t)V_{q}(t)\;,
\label{linear_tracelog}
\end{align}
where we suppressed the spatial index for the sake of compactness, and we used that $G_0^A(t,t)+G_0^R(t,t)$ must be zero due to the causality structure. The latter properties ensures as well that $Z|_{V_{q}=0}=1$.
The second term reads
\begin{align}
\label{rpa_second}
&\mathrm{Tr}\left[\underline{\mathbf{G}}_0\cdot\underline{\mathbf{V}}\cdot \underline{\mathbf{G}}_0\cdot\underline{\mathbf{V}}\right]=\int_{-\infty}^\infty\!\!dtdt^\prime \left[G_0^A(t,t^\prime)V_{cl}(t^\prime)G_0^A(t^\prime,t)V_{cl}(t)+G_0^R(t,t^\prime)V_{cl}(t^\prime)G_0^R(t^\prime,t)V_{cl}(t)\right]\nonumber\\
+&\int_{-\infty}^\infty\!\!dtdt^\prime \left[G_0^A(t,t^\prime)V_{q}(t^\prime)G_0^R(t^\prime,t)V_{q}(t)+G_0^R(t,t^\prime)V_{q}(t^\prime)G_0^A(t^\prime,t)V_{q}(t)+G_0^K(t,t^\prime)V_{q}(t^\prime)G_0^K(t^\prime,t)V_{q}(t)\right]\nonumber\\
+&\int_{-\infty}^\infty\!\!dtdt^\prime \left[G_0^R(t,t^\prime)V_{q}(t^\prime)G_0^K(t^\prime,t)V_{cl}(t)+G_0^K(t,t^\prime)V_{q}(t^\prime)G_0^A(t^\prime,t)V_{cl}(t)\right]\nonumber\\
+&\int_{-\infty}^\infty\!\!dtdt^\prime \left[G_0^A(t,t^\prime)V_{q}(t^\prime)G_0^K(t^\prime,t)V_{cl}(t)+G_0^K(t,t^\prime)V_{q}(t^\prime)G_0^R(t^\prime,t)V_{cl}(t)\right]\;.
\end{align}
Now, the first line in~(\ref{rpa_second}) is zero again for the same reason for which $Z|_{V_{q}=0}=1$. The following three lines contain instead the $(R,A,K)$ components of the polarization propagator, $\Pi^K,\Pi^R$ and $\Pi^A$, as we shall see below.

In order to derive explicit expressions for the $(R,A,K)$ components of the effective cavity propagator, we go back to the momentum-frequency representation given in Eqs.~(\ref{atom_rak_explicit}),(\ref{sources_explicit}).
The term linear in $V$ reads:
\begin{align}
\label{disp_shift}
i\mathrm{Tr}\left[\underline{\mathbf{G}}_0\cdot\underline{\mathbf{V}}\right]&=i\sum_{\substack{\mathbf{k},\mathbf{k}^\prime}}\int_{-\infty}^\infty\frac{d\omega d\omega^\prime}{(2\pi)^2}G_0^K(\omega,\mathbf{k})\delta_{\mathbf{k},\mathbf{k}^\prime}2\pi\delta(\omega-\omega^\prime)V_q(\omega-\omega^\prime,\mathbf{k}-\mathbf{k}^\prime)\nonumber\\
&=i\sum_{\substack{\mathbf{k}}}\int_{-\infty}^\infty\frac{d\omega}{2\pi}G_0^K(\omega,\mathbf{k})\frac12\frac{g_0^2}{\Delta_{\rm a}}\frac12\int_{-\infty}^\infty\frac{d\omega^\prime}{2\pi}\left(a_{cl}^*(\omega^\prime) a_{q}(\omega^\prime)+a_{q}^*(\omega^\prime) a_{cl}(\omega^\prime)\right)=\nonumber\\
&=\frac12\frac{g_0^2}{\Delta_{\rm a}}\underbrace{\sum_{\substack{\mathbf{k}}}F(\epsilon_{\mathbf{k}})}_{=\infty-2N}\frac12\int_{-\infty}^\infty\frac{d\omega^\prime}{2\pi}\left(a_{cl}^*(\omega^\prime) a_{q}(\omega^\prime)+a_{q}^*(\omega^\prime)a_{cl}(\omega^\prime)\right)\nonumber\\
&=-\frac12\frac{g_0^2}{\Delta_{\rm a}}N\int_{-\infty}^\infty\frac{d\omega}{2\pi}\left(a_{cl}^*(\omega) a_{q}(\omega)+a_{q}^*(\omega)a_{cl}(\omega)\right)\;,
\end{align}
where we used the equilibrium distribution function (\ref{dist_eq}) and neglected the irrelevant divergence resulting from the sum over $\mathbf{k}$ of one. This infinity is rightfully neglected since it is a spurious result of the continuous limit of the atomic propagator and is therefore absent in the correct discrete form. Note that the pump-cavity scattering does not contribute at this level due to the fully non diagonal geometrical factor $\eta_{\rm PC}(\mathbf{k}-\mathbf{k}^\prime)$. The term (\ref{disp_shift}) corresponds to the usual dispersive shift of the bare cavity frequency due to the two-level atoms, so that $-\Delta_{\rm c}\to -\Delta_{\rm c}+g_0^2N/2\Delta_{\rm a}$.

For the term quadratic in $V$ we decompose
\begin{equation}
\label{quadratic_effective_action}
\frac{i}{2}\mathrm{Tr}\left[\underline{\mathbf{G}}_0\cdot\underline{\mathbf{V}}\cdot \underline{\mathbf{G}}_0\cdot\underline{\mathbf{V}}\right]=S^{(2),K}+S^{(2),R}+S^{(2),A}\;.
\end{equation}
Using the compact notation $(\omega,\mathbf{k})\to(1),(\omega^\prime,\mathbf{k}^\prime)\to(2)$, the Keldysh part thus reads
\begin{align}
S^{(2),K}&=\frac{i}{2}\int d(1)d(2)V_q(1-2)V_q(2-1)\left[G_0^A(1) G_0^R(2)+G_0^R(1) G_0^A(2)+G_0^K(1) G_0^K(2)\right]\\
&=\frac{i}{2}\int d(1)d(2)V_q(1-2)V_q(2-1)\left[G_0^K(1) G_0^K(2)-(G_0^R(1)- G_0^A(1))(G_0^R(2)- G_0^A(2))\right]\nonumber\;, 
\end{align}
where the second equality comes due to the fact that $G^{R(A)}(t,t^\prime)G^{R(A)}(t^\prime,t)=0$. Since the atoms are assumed to be described by an equilibrium scalar theory, we have $G_0^K(1)=F(1)\left(G_0^R(1)-G_0^A(1)\right)$, with $F(1)=\tanh((\omega-\mu)/2T)$, so that 
\begin{align}
&S^{(2),K}=\frac{i}{2}\int d(1)d(2)V_q(1-2)V_q(2-1)\left[\left(G_0^R(1)-G_0^A(1)\right) \left(G_0^R(2)-G_0^A(2)\right) \underbrace{\left(F(1)F(2)-1\right)}_{F^{-1}(1-2)(F(2)-F(1))}\right]\nonumber\\ 
&=\frac{i}{2}\int d(1)d(2)V_q(1-2)V_q(2-1)F^{-1}(1-2)\left[G_0^K(2)\left(G_0^R(1)-G_0^A(1)\right)-G_0^K(1)\left(G_0^R(2)-G_0^A(2)\right)\right]\nonumber\\ 
&=\frac{i}{2}\int d(1)d(2)V_q(1)V_q(-1)F^{-1}(1)\big|_{\mu=0}\Bigg[G_0^R(1+2)G_0^K(2)+G_0^K(1+2)G_0^A(2)
\nonumber\\
&-\left(G_0^A(1+2)G_0^K(2)-G_0^K(1+2)G_0^R(2)\right)\Bigg]\nonumber\;,
\end{align}
where $F(1-2)=\tanh((\omega-\omega^\prime)/2T)$ does not depend on $\mu$ anymore. Now, upon defining
\begin{align}
\Pi^{R(A)}(1)&=\frac{i}{2}\int d(2)\left[G_0^{R(A)}(1+2)G_0^K(2)+G_0^K(1+2)G_0^{A(R)}(2)\right]\\
\Pi^K(1)&=F(1)\big|_{\mu=0}\left[\Pi^R(1)-\Pi^A(1)\right]\;,
\end{align}
where the last line expresses the \emph{bosonic} fluctuation-dissipation theorem (FDT), we have
\begin{align}
S^{(2),K}=\int d(1)\;\Pi^K(1)V_q(-1)V_q(1)\;.
\end{align}
For the advanced and retarded parts of the quadratic effective action we get similarly
\begin{align}
S^{(2),R}&=\int d(1)\;\Pi^R(1)V_{cl}(-1)V_q(1)\\
S^{(2),A}&=\int d(1)\;\Pi^A(1)V_q(-1)V_{cl}(1)\;.
\end{align}
The explicit forms of the components of the polarization propagator read

\begin{align}
\Pi^R(\omega,\mathbf{k})&=\frac{i}{2}\sum_{\substack{\mathbf{k}^\prime}}\int_{-\infty}^\infty\frac{d\omega^\prime}{2\pi}\left[G_0^R(\omega+\omega^\prime,\mathbf{k}+\mathbf{k}^\prime) G_0^K(\omega^\prime,\mathbf{k}^\prime)+G_0^K(\omega+\omega^\prime,\mathbf{k}+\mathbf{k}^\prime) G_0^A(\omega^\prime,\mathbf{k}^\prime)\right]\nonumber\\
&=\frac12 \sum_{\substack{\mathbf{k}^\prime}}\left[\frac{F(\epsilon_{\mathbf{k}^\prime})}{\omega+\epsilon_{\mathbf{k}^\prime}-\epsilon_{\mathbf{k}+\mathbf{k}^\prime}+i0^+}+\frac{F(\epsilon_{\mathbf{k}+\mathbf{k}^\prime})}{\epsilon_{\mathbf{k}+\mathbf{k}^\prime}-\omega-\epsilon_{\mathbf{k}^\prime}-i0^+}\right]\nonumber\\
&=\sum_{\substack{\mathbf{k}^\prime}}\frac{n_{\rm F}(\epsilon_{\mathbf{k}+\mathbf{k}^\prime})-n_{\rm F}(\epsilon_{\mathbf{k}^\prime})}{\omega+\epsilon_{\mathbf{k}^\prime}-\epsilon_{\mathbf{k}+\mathbf{k}^\prime}+i0^+}\;,\label{polariz_ret}
\end{align}
$\Pi^A(\omega,\mathbf{k})=\Pi^R(\omega,\mathbf{k})^*$, 
and finally the Keldysh component
\begin{align}
\Pi^K(\omega,\mathbf{k})&=F^{-1}(\omega)\big|_{\mu=0}2i\mathrm{Im}\Pi^R(\omega,\mathbf{k})
\;.
\end{align}
Up to order one in the $1/N$ expansion, we consider only contributions up to second order in the cavity field $a_{cl,q}$, so that in Eq.~(\ref{quadratic_effective_action}) we have to take only the terms linear in $a_{cl,q}$ in each $V$. Eq.~(\ref{quadratic_effective_action}) would thus contain terms involving $\sum_{\mathbf{k}}\eta_{\rm PC}(\mathbf{k})\eta_{\rm PC}(-\mathbf{k})\Pi^{R,A,K}(\omega,\mathbf{k})$. Since both $\eta_{\rm PC}(\mathbf{k})$ and $\Pi^{R,A,K}(\omega,\mathbf{k})$ are symmetric in $\mathbf{k}\to-\mathbf{k}$, and assuming $\mathbf{Q}$ to be the momentum transfer involved in the two-photon transition, we will get 
\[
\sum_{\mathbf{k}}\eta_{\rm PC}(\mathbf{k})\eta_{\rm PC}(-\mathbf{k})\Pi^{R,A,K}(\omega,\mathbf{k})=\left\{\begin{array}{c}\Pi^{R,A,K}(\omega,\mathbf{Q})/2,\;\;\text{d=1}\\\Pi^{R,A,K}(\omega,\mathbf{Q})/4,\;\;\text{d=2}\end{array}\right.\;,
\]
where $\mathbf{Q}=\mathbf{Q}_{\rm c}$ in $d=1$ and $\mathbf{Q}=\mathbf{Q}_{\rm c}+\mathbf{Q}_{\rm p}$ in $d=2$.



\end{document}